\documentclass[12pt]{article}
\pdfoutput=1
\usepackage{amsmath,amssymb}
\usepackage{bm}
\usepackage{graphicx}
\usepackage{color}
\usepackage{wrapfig}
\begin{document}
\title{
\begin{flushright}
\ \\*[-80pt]
\begin{minipage}{0.17\linewidth}
\normalsize
HUPD1709 \\ \\ \\
\end{minipage}
\end{flushright}
{\Large \bf Towards the minimal seesaw model \\ via CP violation of neutrinos
\\*[20pt]}}

\author{
\centerline{Yusuke~Shimizu$^{1,}$\footnote{E-mail address: yu-shimizu@hiroshima-u.ac.jp} \ ,
 Kenta~Takagi$^{1,}$\footnote{E-mail address: takagi-kenta@hiroshima-u.ac.jp}  \ ,  and
Morimitsu~Tanimoto$^{2,}$\footnote{E-mail address: tanimoto@muse.sc.niigata-u.ac.jp} }
\\*[20pt]
\centerline{
\begin{minipage}{\linewidth}
\begin{center}
$^1${\it \normalsize
\it \normalsize \it \normalsize Graduate School of Science, Hiroshima University, \\ Higashi -Hiroshima 739-8526, Japan}
\\*[4pt]
$^2${\it \normalsize
Department of Physics, Niigata University,~Niigata 950-2181, Japan }
\end{center}
\end{minipage}}
\\*[70pt]}

\date{
\centerline{\small \bf Abstract}
\begin{minipage}{0.9\linewidth}
\vskip  1 cm
\small 
We study  the minimal seesaw model,
where two right-handed Majorana neutrinos are introduced,
focusing on the CP violating phase.
In addition, we take the trimaximal mixing pattern for the neutrino flavor where the charged lepton mass matrix is diagonal.
Owing to this symmetric  framework,
the $3\times 2$ Dirac neutrino mass matrix is given in terms of a few parameters.
It is found that
the observation of the CP violating phase  determines the flavor structure of the Dirac neutrino mass matrix
in the minimal seesaw model.
New  minimal Dirac neutrino mass matrices are presented
in the case of $\rm TM_1$, which is given by the additional 2-3 family mixing to the tri-bimaximal mixing basis in the normal hierarchy of neutrino masses.
Our model includes the Littlest seesaw model by King {\it et al.}
as one of the specific cases.
Furthermore, it is remarked that
our $3\times 2$ Dirac neutrino mass matrix is reproduced by introducing gauge singlet flavons with the specific alignments of the VEV's.
These alignments are derived from the residual symmetry of  $S_4$ group.
\end{minipage}
}

\begin{titlepage}
\maketitle
\thispagestyle{empty}
\end{titlepage}

\section{Introduction}

The standard model (SM) has been well established by the discovery of the Higgs boson.
However, the origin of flavor in quarks and leptons is still unknown
in spite of the remarkable success of the SM.
The underlying physics for the flavor in quarks and leptons is one of the fundamental problems in particle physics.

On the other hand, the neutrino oscillation experiments are going on a new step to observe the CP violation in the lepton sector.
The T2K experiment has confirmed the neutrino oscillation in the $\nu_\mu\to\nu_e$ appearance events \cite{Abe:2013hdq},
which provides us a new information of the CP violation in the lepton sector.
Recent T2K data strongly indicate the CP violation in the neutrino oscillation \cite{Abe:2017vif,T2K}.
The NO$\nu$A data also suggest the CP violation \cite{Adamson:2017gxd},
which is consistent with the T2K result.
Since  the experimental data of the CP violating phase will be available
in the near future,
 one can develop the flavor structure of Yukawa couplings by taking account of it.

We study the flavor structure in the seesaw model \cite{Minkowski}-\cite{MSSV}
in order to find a clue of  the underlying physics of flavor.
It is advantageous to consider the minimum number of parameters needed for reproducing the neutrino mixing angles and CP violating phases completely \cite{Shimizu:2012ry}.
Indeed, there are some attempts toward so-called the minimal seesaw model \cite{King:1998jw}-\cite{Rink:2016knw}.

In this work, we investigate the minimal seesaw model via the CP violation,
where we assume two right-handed Majorana neutrinos.
In addition, we take the trimaximal mixing pattern ~\cite{Grimus:2008tt,Albright:2008rp} of the neutrino flavor
  to reduce the number of free parameters of the Dirac neutrino mass matrix.
 The trimaximal mixing pattern is derived from the flavor symmetry.

Before the reactor experiments reported the non-zero value of $\theta_{13}$ in 2012 \cite{An:2012eh,Ahn:2012nd},
there has been a paradigm of the tri-bimaximal (TBM) mixing ~\cite{Harrison:2002er,Harrison:2002kp}.
Since the TBM mixing pattern  is  a highly symmetric,
the non-Abelian discrete groups have become the center of attention at the flavor symmetry
\cite{Ma:2001dn}-\cite{King:2014nza}.
The observation of the non-vanishing $\theta_{13}$ forces to study the  deviation from the  TBM mixing.  It is remarked that the trimaximal mixing pattern is still available  after the observation of $\theta_{13}$.
The trimaximal mixing is  given by the additional rotation of 2-3 ($\rm TM_1$) or 1-3 ($\rm TM_2$) families of neutrinos
to the TBM mixing basis~\cite{Albright:2010ap,Rodejohann:2012cf}.
The additional   rotation of 1-2 families is called as $\rm TM_3$, 
 however it still leads to
$\theta _{13}=0$. We do not discuss the additional rotation of 1-2 families, $\rm TM_3$, since it leads to $\theta _{13}=0$.

One obtains testable relations among the neutrino mixing angles and  the CP violating phase,  so-called mixing sum rules in $\rm TM_1$ and $\rm TM_2$ \cite{Antusch:2011ic}-\cite{Petcov:2014laa}.
The prediction for the CP violation of neutrinos has a big impact on the study of the flavor structure \cite{Marzocca:2013cr}-\cite{Morozumi:2017rrg}
because T2K and NO$\nu$A experiments are expected to confirm the non-zero CP violating phase in the near future.
We discuss  the CP violation of neutrinos for both $\rm TM_1$ and $\rm TM_2$ in the framework of the minimal seesaw model to restrict the structure of the Dirac neutrino mass matrix.
It is found that
those Dirac neutrino mass matrices are reproduced by introducing gauge singlet flavons with the vacuum expectation value (VEV) in the non-Abelian discrete symmetry,
for example,  $S_4$ symmetry.
The specific alignments of the VEV's suggest the residual symmetry of  
 $S_4$ group in our work.

We investigate   the Dirac neutrino mass matrix in the diagonal basis of the charged lepton mass matrix and the $2\times 2$ right-handed Majorana neutrino mass matrix
focusing the CP violating Dirac phase.
Since our results depend on the neutrino mass hierarchies,
we discuss both cases of the normal hierarchy (NH) and the inverted hierarchy (IH) for neutrino masses, respectively.

Our analyses include the Littlest seesaw model 
by King {\it et al.}~\cite{King:2015dvf}-\cite{King:2016yvg}
as the  one of the specific cases of our model.


The paper is organized as follows.
We present our framework of the minimal seesaw model in section 2,
where the structure of the Dirac neutrino mass matrix is discussed to reproduce $\rm TM_1$ and $\rm TM_2$ for both cases of  NH and IH.
The Dirac neutrino mass matrix is also discussed in the view of the flavor symmetry.
In section 3, the numerical results are presented.
The section 4 is devoted to the summary and discussions.
Appendix A gives the detail studies of the Dirac neutrino mass matrix,
and Appendix B presents the necessary group theory of $S_4$.

\section{Realization of minimal  seesaw model}
Let us start with discussing the structure of the lepton mixing matrix,
so-called the Pontecorvo-Maki-Nakagawa-Sakata (PMNS) matrix $U_{\text{PMNS}}$~\cite{Maki:1962mu,Pontecorvo:1967fh},
where neutrinos are supposed to be Majorana particles.
It is parametrized in terms of three mixing angles $\theta _{ij}$ $(i,j=1,2,3;~i<j)$,
one CP violating Dirac phase $\delta _{CP}$, and two Majorana phases $\alpha,\ \beta$ as follows:
\begin{align}
U_\text{PMNS} \equiv
\begin{pmatrix}
c_{12} c_{13} & s_{12} c_{13} & s_{13}e^{-i\delta _{CP}} \\
-s_{12} c_{23} - c_{12} s_{23} s_{13}e^{i\delta _{CP}} &
c_{12} c_{23} - s_{12} s_{23} s_{13}e^{i\delta _{CP}} & s_{23} c_{13} \\
s_{12} s_{23} - c_{12} c_{23} s_{13}e^{i\delta _{CP}} &
-c_{12} s_{23} - s_{12} c_{23} s_{13}e^{i\delta _{CP}} & c_{23} c_{13}
\end{pmatrix}
\times
\begin{pmatrix}
e^{i\frac{\alpha}{2}}&0 &0 \\
0 & e^{i\frac{\beta}{2}} & 0 \\
0 & 0 & 1
\end{pmatrix},
\label{UPMNS}
\end{align}
where $c_{ij}$ and $s_{ij}$ denote $\cos\theta_{ij}$ and $\sin\theta_{ij}$, respectively.
The CP violating measure, Jarlskog invariant~\cite{Jarlskog:1985ht},
is defined by the PMNS matrix elements $U_{\alpha i}$,
and  is written in terms of the mixing angles and the CP violating phase as:
\begin{equation}
J_{CP}=\text{Im}\left [U_{e1}U_{\mu 2}U_{e2}^\ast U_{\mu 1}^\ast \right ]
=s_{23}c_{23}s_{12}c_{12}s_{13}c_{13}^2\sin \delta _{CP}~.
\label{Jcp}
\end{equation}

For the lepton mixing matrix, Harrison-Perkins-Scott proposed a simple form of the mixing matrix,
so-called TBM mixing ~\cite{Harrison:2002er,Harrison:2002kp} as follows:
\begin{equation}
V_{\text{TBM}}=
\begin{pmatrix}
\frac{2}{\sqrt{6}} & \frac{1}{\sqrt{3}} & 0 \\
-\frac{1}{\sqrt{6}} & \frac{1}{\sqrt{3}} & -\frac{1}{\sqrt{2}} \\
-\frac{1}{\sqrt{6}} & \frac{1}{\sqrt{3}} & \frac{1}{\sqrt{2}}
\end{pmatrix},
\label{UTBM}
\end{equation}
which was a good scheme for the lepton sector
before the reactor experiments reported non-zero $\theta _{13}$.
Therefore, it is reasonable to take the TBM mixing  as the starting point of the lepton mixing.
In order to avoid vanishing  $\theta _{13}$,
 we move to the trimaximal mixing basis from the TBM mixing  basis.

The minimal seesaw model consists of two right-handed Majorana neutrinos
and three left-handed neutrinos in Type I seesaw \cite{Shimizu:2012ry}.
Taking both the charged lepton mass matrix and right-handed Majorana neutrino one $M_R$ to be real diagonal,
$M_R$ and the Dirac neutrino mass matrix $M_D$ are generally written as:
\begin{equation}
M_R=-M_0
\begin{pmatrix}
p^{-1}
 & 0 \\
0 & 1
\end{pmatrix}, \qquad\qquad
M_D=
\begin{pmatrix}
a & d \\
b & e \\
c & f
\end{pmatrix},
\end{equation}
respectively,
where $a\sim f$ are complex parameters 
and $p$ is the ratio between the two right-handed Majorana neutrino masses.
The minus sign in front of $M_0$ is taken as our sign convention.
By using the seesaw mechanism of Type I, the left-handed Majorana neutrino mass matrix $M_{\nu }$ is give by
\begin{equation}
M_{\nu }=-M_DM_R^{-1}M_D^T=\frac{1}{M_0}
\begin{pmatrix}
a^2p+d^2 & abp+de & acp+df \\
abp+de & b^2p+e^2 & bcp+ef \\
acp+df & bcp+ef & c^2p+f^2
\end{pmatrix}.
\label{left-handed-Majorana-1}
\end{equation}
By turning the neutrino mass matrix $M_\nu $ to the TBM mixing  basis,
the left-handed Majorana neutrino mass matrix is given as
\begin{align}
\hat M_\nu\equiv  V_{\text{TBM}}^T M_\nu V_{\text{TBM}}=\frac{1}{M_0}
\begin{pmatrix}
\frac{A_\nu ^2p+D_\nu^2}{6} & \frac{A_\nu B_\nu p+D_\nu  E_\nu}{3\sqrt{2}} &
\frac{A_\nu C_\nu p+D_\nu F_\nu}{2\sqrt{3}} \\
\frac{A_\nu B_\nu p+D_\nu  E_\nu}{3\sqrt{2}} & \frac{B_\nu^2p+E_\nu^2}{3} &
 \frac{B_\nu C_\nu p+E_\nu F_\nu}{\sqrt{6}} \\
\frac{A_\nu C_\nu p+D_\nu F_\nu}{2\sqrt{3}} &\frac{B_\nu C_\nu p+E_\nu F_\nu}{\sqrt{6}} & \frac{C_\nu^2p+F_\nu^2}{2}
\end{pmatrix},
\label{left-handed-Majorana-TBM}
\end{align}
where
\begin{align}
&& A_\nu\equiv 2a-b-c, \qquad B_\nu\equiv a+b+c, \qquad C_\nu\equiv c-b, \nonumber\\
&& D_\nu\equiv 2d-e-f, \qquad E_\nu\equiv d+e+f, \qquad F_\nu\equiv f-e.
\end{align}
The mass matrix $\hat M_\nu$ is  the left-handed Majorana neutrino mass matrix in the TBM mixing basis.
In Appendix A, we have classified this neutrino mass matrix to reproduce the trimaximal mixing $\rm TM_1$ and $\rm TM_2$ by the additional rotation of  2-3 or 1-3 families for both cases of  NH and IH.

\subsection{$\rm TM_1$: 2-3 family mixing in NH}
First of all, we consider the case of $\rm TM_1$  in NH.
The non-vanishing $\theta_{13}$ is obtained 
by an additional 2-3 family mixing to the TBM mixing  basis  ~\cite{Shimizu:2012ry,Rodejohann:2012cf}.
This case is given by the following Dirac neutrino mass matrix as shown in Appendix A.1:
 \begin{equation}
 M_D=
 \begin{pmatrix}
 \frac{b+c}{2} & \frac{e+f}{2} \\
 b & e \\
 c & f
 \end{pmatrix},
 \label{MDgeneral}
 \end{equation}
which leads to the neutrino mass matrix
\begin{equation}
\hat M_\nu =\frac{f^2}{M_0}
\begin{pmatrix}
0 & 0 & 0 \\
0 & \frac{3}{4}\left [B^2 e^{2i\phi_B}(1+j)^2+(k+1)^2\right ] &
\frac{1}{2}\sqrt{\frac{3}{2}}\left [B^2 e^{2i\phi_B}(1-j^2)-k^2+1\right ]\\
0 & \frac{1}{2}\sqrt{\frac{3}{2}}\left [B^2 e^{2i\phi_B}(1-j^2)-k^2+1\right ] &
\frac{1}{2}\left [B^2 e^{2i\phi_B}(1-j)^2+(k-1)^2\right ]
\end{pmatrix},
\label{left-handed-1}
\end{equation}
where we put $p=1$ in Eq.~(\ref{left-handed-Majorana-TBM}) after rescaling parameters.
This matrix is derived from Eq.~(\ref{Aleft-handed-Majorana-general}) 
in Appendix A,
where we can take $e$ and $f$ to be real, and $b$ and $c$ to be complex in general by using the freedom of redefinitions of phases in the left-handed lepton fields.
We consider the case that  the relative phase between $b$ and $c$ is $0$ or $\pi$ in order to reduce the number of parameters of the neutrino mass matrix.
Then, the only one phase which leads to the CP violation of neutrinos
is the relative phase between the first and the second columns of the Dirac neutrino mass matrix in Eq.~(\ref{MDgeneral}), that is the phase of $c/f$.
Since we neglect  the relative phase between $b$ and $c$,
 the effect of it  is discussed in subsection 3.2.
Parameters in Eq.~(\ref{left-handed-1}) are defined as
\begin{equation}
 \frac{e}{f}=k \ , \qquad\qquad  \frac{b}{c}=j  \ , \qquad\qquad
  \frac{c}{f}=B e^{i\phi_B}\ ,
\label{parameters}
\end{equation}
where $k$, $j$, and $B$ are real.

The neutrino mass matrix in Eq.~(\ref{left-handed-1}) is diagonalized by the rotation of 2-3 families as
\begin{equation}
V_{23}=\frac{1}{{\cal A}}
\begin{pmatrix}
{\cal A} & 0 & 0 \\
0 & 1& {\cal V}\\
0 &  -{\cal V}^* & 1
\end{pmatrix}, \qquad\qquad {\cal A}=\sqrt{1+|{\cal V}|^2} \ ,
\label{rotation23}
\end{equation}
where ${\cal V}$ is given in terms of $k$, $j$, $B$, and $\phi_B$.
Therefore, the PMNS matrix is calculated as follows:
\begin{align}
U_\text{PMNS}= V_{\text{TBM}} V_{23} \ ,
\end{align}
which gives three mixing angles, one Dirac phase, and one Majorana phase.

On the other hand, three neutrino masses are related to those parameters as follows:
\begin{eqnarray}
&& m_1=0, \qquad\qquad m_2^2 m_3^2=\frac{9}{4}(j-k)^4 B^4 f^8 \ ,  \\
&&m_2^2+m_3^2=\frac{f^4}{16}\left [B^4(5j^2+2j+5)^2+
2B^2(5jk+j+k+5)^2\cos 2\phi_B+(5k^2+2k+5)^2
\right ],\nonumber
\label{mass-0}
\end{eqnarray}
which indicate that the PMNS matrix elements are correlated with neutrino masses.

At first, we study the specific Dirac neutrino mass matrices where
 the relative phase of $b/c$ vanishes. These cases lead to  one zero textures
 for the Dirac neutrino mass matrix (see Appendix A).
For case I, we consider the case of  $j=b/c=-1$. 
The neutrino mass matrix is given as 
\begin{equation}
{\rm Case\ I}\ : \
\hat M_{\nu }=\frac{f^2}{M_0}
\begin{pmatrix}
0 & 0 & 0 \\
0 & \frac{3}{4}(k+1)^2 & -\frac{1}{2}\sqrt{\frac{3}{2}}(k^2-1) \\
0 & -\frac{1}{2}\sqrt{\frac{3}{2}}(k^2-1) & 2B^2  e^{2i\phi_B}+\frac{1}{2}(k-1)^2
\end{pmatrix},
\label{case1}
\end{equation}
where the Dirac neutrino mass matrix is 
 \begin{equation}
M_D=
\begin{pmatrix}
0 & \frac{e+f}{2} \\
b & e \\
-b & f
\end{pmatrix}.
\label{MD1}
\end{equation}

For case II where  $c=0$ is put, it is
\begin{equation}
{\rm Case\ II}\ : \
\hat M_{\nu }=\frac{f^2}{M_0}
\begin{pmatrix}
0 & 0 & 0 \\
0 & \frac{3}{4}[\hat B^2 e^{2i\phi_B}+(k+1)^2] &
 -\frac{1}{2}\sqrt{\frac{3}{2}}[\hat B^2 e^{2i\phi_B}+k^2-1] \\
0 &  -\frac{1}{2}\sqrt{\frac{3}{2}}[\hat B^2 e^{2i\phi_B}+k^2-1] &
\frac{1}{2}[\hat B^2 e^{2i\phi_B}+(k-1)^2]
\end{pmatrix}
,
\end{equation}
which corresponds to $j=\infty$ and $B=0$ with a finite quantity $\hat B \equiv Bj$, that is
\begin{equation}
M_D=
\begin{pmatrix}
\frac{b}{2} & \frac{e+f}{2} \\
b & e \\
0 & f
\end{pmatrix}.
\label{MD2}
\end{equation}

For case III where  $b=0$,  it is

\begin{equation}
{\rm Case\ III}\ : \
\hat M_{\nu }=\frac{f^2}{M_0}
\begin{pmatrix}
0 & 0 & 0 \\
0 & \frac{3}{4}[B^2 e^{2i\phi_B}+(k+1)^2] &
-\frac{1}{2}\sqrt{\frac{3}{2}}[-B^2 e^{2i\phi_B}+k^2-1] \\
0 &  -\frac{1}{2}\sqrt{\frac{3}{2}}[-B^2 e^{2i\phi_B}+k^2-1] &
 \frac{1}{2}[B^2 e^{2i\phi_B}+(k-1)^2]
\end{pmatrix}
,
\end{equation}
which  corresponds to $j=0$, that is
\begin{equation}
M_D=
\begin{pmatrix}
\frac{c}{2} & \frac{e+f}{2} \\
0 & e \\
c & f
\end{pmatrix}.
\label{MD3}
\end{equation}

It is interesting   to note  the relation between our neutrino mass matrices and the Littlest seesaw model by King {\it et al.}~\cite{King:2015dvf}-\cite{King:2016yvg}.
It corresponds to $k=-3$ in case I, that is,
\begin{equation}
M_D=
\begin{pmatrix}
0 & f \\
b & 3f \\
-b & -f
\end{pmatrix}.
\label{King}
\end{equation}
We discuss the phenomenological  implication for this model.

\subsection{$\rm TM_1$: 2-3 family mixing in IH}

Next, we discuss the IH case of neutrino masses in $\rm TM_1$.
As shown in Appendix A.2, the Dirac and left-handed Majorana neutrino mass matrices are given as:
\begin{equation}
 M_D=
\begin{pmatrix}
-2b & \frac{e+f}{2} \\
b & e \\
b & f
\end{pmatrix},
\label{MD23-IH}
\end{equation}
and
\begin{equation}
\hat M_{\nu }=\frac{1}{M_0}
\begin{pmatrix}
6b^2 & 0 & 0 \\
0 & \frac{3}{4}(e+f)^2 & -\frac{1}{2}\sqrt{\frac{3}{2}}(e-f)(e+f) \\
0 & -\frac{1}{2}\sqrt{\frac{3}{2}}(e-f)(e+f) & \frac{1}{2}(e-f)^2
\end{pmatrix}
,
\end{equation}
where $m_1=6b^2/M_0$ and $m_3=0$.
By taking the elements in the first column of the Dirac neutrino mass matrix to be real,
parameters $e$ and $f$ can be complex in contrast to the case of subsection 2.1.
When we redefine the complex parameters $e$ and $f$ in terms of two real parameters $k$ and $\phi_k$ as
\begin{equation}
\frac{e}{f}=k e^{i\phi_k} \ ,
\label{parametersIH}
\end{equation}
the neutrino mass matrix turns to
 \begin{equation}
 \hat M_{\nu }=\frac{6b^2}{M_0}
 \begin{pmatrix}
1 & 0 & 0 \\
 0 & 0 & 0\\
 0 & 0 & 0
 \end{pmatrix} + \frac{f^2}{M_0}
  \begin{pmatrix}
 0 & 0 & 0 \\
 0 & \frac{3}{4}(ke^{i\phi_k}+1)^2 & -\frac{1}{2}\sqrt{\frac{3}{2}}(k^2e^{2i\phi_k}-1) \\
 0 & -\frac{1}{2}\sqrt{\frac{3}{2}}(k^2e^{2i\phi_k}-1) & \frac{1}{2}(ke^{i\phi_k}-1)^2
 \end{pmatrix}
 .  \label{Mnu23-IH}
 \end{equation}
Note that the $2$-$3$ family mixing is determined only by $k$ and $\phi_k$.
If the relative phase between $e$ and $f$, that is $\phi_k$, is zero,
the  CP symmetry is conserved.
It is remarked that the 2-3 family mixing is independent of neutrino masses $m_1$ and $m_2$
because the neutrino masses $m_1$ and  $m_2$ are given in terms of $b$ and   $f$,
respectively.
This situation makes our numerical analysis simple for the case of IH.

Besides our neutrino mass matrix,  
the two-zero texture for the Dirac neutrino mass matrix
has been discussed in the  context of the minimal seesaw model
\cite{Frampton:2002qc,Harigaya:2012bw,Zhang:2015tea,Rink:2016vvl}.
Especially, for the IH case, 
it is completely consistent with the experimental data of mixing angles and masses
 \cite{Harigaya:2012bw}.
The prediction of the CP violating phase
 will be discussed comparing with our result in section 3.

\subsection{$\rm TM_2$: 1-3 family mixing in NH or IH}

Let us consider the other case, in which the additional rotation of $1$-$3$ families diagonalizes the neutrino mass matrix $\hat M_\nu$.
As seen in Appendix A.3,
we obtain the Dirac neutrino mass matrix and the left-handed Majorana neutrino mass matrix as follows:
\begin{equation}
 M_D=
\begin{pmatrix}
b & -e-f \\
b & e \\
b& f
\end{pmatrix},
\label{MD13}
\end{equation}
\begin{equation}
\hat M_\nu =\frac{1}{M_0}
\begin{pmatrix}
\frac{3}{2}(e+f)^2 & 0 &\frac{\sqrt{3}}{2}(e^2-f^2)\\
0 & 3b^2 & 0 \\
\frac{\sqrt{3}}{2}(e^2-f^2) &0 & \frac{1}{2}(e-f)^2
\end{pmatrix},
\label{Mnu13}
\end{equation}
respectively.
By using the same notation in Eq.~(\ref{parametersIH}),
$\hat M_\nu$ is  rewritten as:
\begin{equation}
\hat M_\nu =\frac{3b^2}{M_0}
\begin{pmatrix}
0 & 0 &0\\
0 & 1& 0 \\
0 &0 & 0
\end{pmatrix}
+\frac{f^2}{M_0}
\begin{pmatrix}
\frac{3}{2}(ke^{i\phi_k}+1)^2 & 0 &\frac{\sqrt{3}}{2}(k^2e^{2i\phi_k}-1)\\
0 & 0 & 0 \\
\frac{\sqrt{3}}{2}(k^2e^{2i\phi_k}-1) &0 & \frac{1}{2}(ke^{i\phi_k}-1)^2
\end{pmatrix}.
\label{Mnu13}
\end{equation}
As well as the case of the 2-3 family mixing for IH in subsection 2.2,
the 1-3 family mixing is also determined only by $k$ and $\phi_k$,
namely, it is independent of neutrino masses.

\subsection{Dirac neutrino mass matrix and vacuum alignment of flavons}

Above specific structures of the $3\times 2$ Dirac neutrino mass matrix are given by introducing flavons
with VEV's in the framework of the non-Abelian discrete symmetry $S_4$.
The Dirac neutrino mass matrices in subsections 2.1, 2.2, and 2.3 can be reproduced by the four types of flavons $\phi_i$.
These four flavons are represented as triplets of   $S_4$ group and have specific alignments of VEV's:
\begin{equation}
\langle\phi_i \rangle \  = \
\begin{pmatrix}
\frac{b+c}{2} \\
c\\
b
\end{pmatrix}\ ,  \quad
\begin{pmatrix}
-2 \\
1\\
1
\end{pmatrix}\ , \quad
\begin{pmatrix}
1 \\
1\\
1
\end{pmatrix}\ , \quad
\begin{pmatrix}
-e-f \\
f\\
e
\end{pmatrix}\ .
\label{VEV1}
\end{equation}
The first two VEV's in Eq.~(\ref{VEV1}) preserve the $SU~(US)$ symmetry
 for ${\bf 3'}$ and ${\bf 3}$, respectively 
 since the generator $SU~(US)$ is expressed as
\begin{equation}
SU=US=\mp \frac{1}{3}
\begin{pmatrix}
-1 & 2 &2\\
2 & 2& -1 \\
2 &-1 & 2
\end{pmatrix} ,
\label{SU}
\end{equation}
for  ${\bf 3}$ and ${\bf 3'}$, respectively.
These VEV's are broken by $S$, $T$, and $U$ (see Appendix B).
Those flavons have the $Z_2$ symmetry with elements $\{ 1, SU\} $.
This $Z_2$ symmetry is a residual one after $S_4$ is broken.
The third VEV in Eq.~(\ref{VEV1}) holds the $S$ symmetry
for both ${\bf 3}$ and ${\bf 3'}$,
but broken by $T$.
The last VEV in Eq.~(\ref{VEV1}) is not preserved by $S$, $T$, nor $U$ unless $e=f$.

Let us reproduce the Dirac neutrino mass matrix in Eq.~(\ref{MDgeneral}) by introducing those flavons.
Suppose that the left-handed lepton $L$ and  the flavons $\phi_{\rm atm}$, 
 $\phi_{\rm sol}$ are $\bf 3'$ of $S_4$,
while the right-handed Majorana neutrinos $\nu_{R1}$ and $\nu_{R2}$ are $\bf 1$ of $S_4$.
The Higgs field $H_u$ is also $\bf 1$.
For the 2-3 family mixing in NH,
the Dirac neutrino mass matrix is reproduced by the Yukawa couplings
\begin{equation}
\frac{y_{\rm atm}}{\Lambda}\phi_{\rm atm} L  H_u\nu_{R1}^c
+\frac{y_{\rm sol}}{\Lambda}\phi_{\rm sol} L  H_u\nu_{R2}^c  \ ,
\label{Yukawa}
\end{equation}
where $y_{\rm atm}$ and $y_{\rm sol}$ are arbitrary coupling constants, $\Lambda$ is
the cut-off scale of  $S_4$ symmetry.
The VEV's of $\phi_{\rm atm}$ and $\phi_{\rm sol}$ are
\begin{equation}
\langle\phi_{\rm atm} \rangle \  \sim \
\begin{pmatrix}
\frac{b+c}{2} \\
c\\
b
\end{pmatrix}\ ,  \qquad
\langle\phi_{\rm sol} \rangle \  \sim \
\begin{pmatrix}
\frac{e+f}{2} \\
f\\
e
\end{pmatrix}\ ,
\label{VEV2}
\end{equation}
since  $S_4$ singlet contraction $\bf 3'\otimes 3'$ implies that 
$L({\bf 3'})\phi({\bf 3'})=L_1\phi_1+L_2\phi_3+L_3\phi_2$ as seen in Appendix B.

For the case of IH of $\rm TM_1$,  the assignments of the irreducible representations should be changed from the one of NH.
We suppose that $L$ and $\phi_{\rm atm}$ are $\bf 3$,
and $\phi_{\rm sol}$ is $\bf 3'$ of $S_4$,
while $\nu_{R1}$ is $\bf 1$ and $\nu_{R2}$ is $\bf 1'$ of $S_4$.
It is easily seen that the Dirac neutrino mass matrix is reproduced by the Yukawa couplings in Eq.~(\ref{Yukawa}),
where the VEV's of $\phi_{\rm atm}$ and $\phi_{\rm sol}$ are
\begin{equation}
\langle\phi_{\rm atm} \rangle \  \sim \
\begin{pmatrix}
-2 \\
1\\
1
\end{pmatrix}\ ,  \qquad
\langle\phi_{\rm sol} \rangle \  \sim \
\begin{pmatrix}
\frac{e+f}{2} \\
f\\
e
\end{pmatrix}\ ,
\label{VEV2}
\end{equation}
since  $\bf 3\otimes 3^{(')}$  also implies that 
$L({\bf 3})\phi({\bf 3^{(')}})=L_1\phi_1+L_2\phi_3+L_3\phi_2$ for $\bf 1^{(')}$.
Thus, the Dirac neutrino mass matrix is reproduced by the relevant assignments of the flavons.
The $2\times 2$ diagonal right-handed Majorana neutrino mass matrix is reproduced
by help of the auxiliary $Z_2$ symmetry in NH while an auxiliary $Z_2$ is not necessary in IH. 
The diagonal charged lepton mass matrix is realized as well as  the Littlest seesaw model~\cite{King:2015dvf} by introducing three $\bf 3$ flavons,
which have VEV's:
  \begin{equation}
  \langle\phi_{e} \rangle \  \sim \
  \begin{pmatrix}
  1 \\
  0\\
  0
  \end{pmatrix}\ ,  \qquad
  \langle\phi_\mu \rangle \  \sim \
  \begin{pmatrix}
  0 \\
 1\\
 0
  \end{pmatrix}\ , \quad
  \langle\phi_\tau \rangle \  \sim \
  \begin{pmatrix}
 0 \\
 0\\
 1
 \end{pmatrix}\ . \quad
  \label{VEV2}
  \end{equation}

For  $\rm TM_2$, that is the case of the  1-3 family mixing,
the Dirac neutrino mass matrix can be reproduced by the relevant Yukawa couplings in the similar way.
However, the situation of symmetry  is different from the case of $\rm TM_1$ because
the fourth VEV in  Eq.(\ref{VEV1}) is not preserved in any subgroups of $S_4$.
In order to obtain desirable Yukawa couplings, we need
an auxiliary $Z_2$ symmetry.

\section{Numerical results}
In this section, we discuss the numerical results for the CP violating Dirac phase $\delta_{CP}$ as well as  neutrino mixing angles.
We also discuss the Majorana phases,
which contribute the effective mass for the neutrinoless double beta decay ($0\nu\beta\beta$ decay).
The neutrino mixing angles are obtained in terms of the PMNS matrix elements $U_{\alpha i}$ of Eq.~\eqref{UPMNS} as follows:
\begin{align}
s_{12}^2\equiv\sin^2\theta_{12}=\frac{|U_{e2}|^2}{1-|U_{e3}|^2}\ ,\quad s_{23}^2\equiv\sin^2\theta_{23}=\frac{|U_{\mu3}|^2}{1-|U_{e3}|^2}\ ,\quad s_{13}^2\equiv\sin^2\theta_{13}=|U_{e3}|^2.
\label{eq:mixings}
\end{align}
The Dirac CP violating phase $\delta_{CP}$ can be calculated by using the Jarlskog invariant in Eq.~\eqref{Jcp}:
\begin{align}
\sin\delta_{CP}&=\frac{J_{CP}}{s_{23}c_{23}s_{12}c_{12}s_{13}c_{13}^2},
\label{eq:sindelta}
\end{align}
where $\cos\delta_{CP}$ is fixed by $|U_{\mu 1}|^2=1/6$ for the case of 2-3 family mixing,
and $|U_{\mu 2}|^2=1/3$ for the case of  1-3 family mixing, respectively.

The effective mass for the $0\nu\beta\beta$ decay is given
in terms of the Dirac phase $\delta_{CP}$ and Majorana phases $\alpha$ and $\beta$ as follows:
\begin{align}
|m_{ee}|&=\left|m_1U_{e1}^2+m_2U_{e2}^2+m_3U_{e3}^2\right|\nonumber\\
&=\left|m_1 c_{13}^2c_{12}^2e^{i\alpha}+m_2 c_{13}^2s_{12}^2e^{i\beta}+m_3 s_{13}^2e^{-2i\delta_{CP}}\right|,\label{eq:4meeNH}
\end{align}
where $m_1=0$ or $m_3=0$ for NH or IH, respectively.
The Majorana phase $\beta$ is directly related to the expression of the effective mass $|m_{ee}|$ for NH,
while $(\alpha-\beta)$ is related to $|m_{ee}|$ for IH.

Let us explain how to obtain our predictions of the CP violation taking the case I of subsection 2.1 ($\rm TM_1$ in NH) as an example.
The result of global analyses is often used as the inputting data to constrain the unknown parameters \cite{Esteban:2016qun,deSalas:2017kay}.
In our calculations, we have adopted the result in Ref.~\cite{deSalas:2017kay}.
At first, by  inputting the  data of  $\Delta m_{13}^2$ and $\Delta m_{12}^2$
 within $3\sigma$ $(1\sigma)$ range in Table 1,
we remove the two free parameters $f^2/M_0$ and $B$ in Eq.~(\ref{case1}).
The remained free parameter $k$ is scanned in the region of $-20\sim 20$ by generating random numbers in the linear scale.
On the other hand, the phase $\phi_B$ is also scanned in the full region of $-\pi\sim \pi$ in the linear scale.
Then, we  calculate three neutrino mixing angles.
These calculated values are judged by using the experimental data within $3\sigma$ ($1\sigma$) range
 as shown in Table 1.
If they are allowed for the experimental data, we keep the point,
in which the CP violating phases and $|m_{ee}|$ are calculated.
Otherwise, we throw the point.
We continue this procedure to obtain enough points for plotting allowed region.


\begin{table}[hbtp]
	\begin{center}
		\begin{tabular}{|c|c|c|}
			\hline
			observable  &\quad \hskip -1.5 cm NH:  \ \   $1 \sigma$ \qquad\qquad  $3\sigma$ \qquad  &
		\ \  \hskip -1.5cm IH:  \quad  $1 \sigma$ \qquad\qquad\ \  $3\sigma$  \\
			\hline
 $|\Delta m_{13}^2|\times10^{3}\ [{\rm eV}^2]$&\ \ \hskip -0.7 cm $2.55\pm 0.04$ ,
 \ \ \    $2.43\sim 2.67$  
			&\  \hskip -0.4 cm $2.49\pm 0.04$  , \quad   $2.37\sim 2.61$   \\
			\hline
 $\Delta m_{12}^2\ \ \times 10^{5}\ [{\rm eV}^2]$&\quad \hskip -0.8 cm$7.56\pm 0.19$ , 	\ \    $7.05\sim 8.14$	&\quad\hskip -0.7 cm $7.56\pm 0.19$  , \ \ \    $7.05\sim 8.14$  \\
			\hline
			$\sin^2\theta_{23}$& \quad\hskip -0.4 cm $0.430^{+0.020}_{-0.018}$\ , \quad $0.384\sim 0.635$ 
			 &\quad\hskip -0.25 cm $0.596^{+0.017}_{-0.018}$  , \quad\  $0.388\sim 0.638$ \\
			\hline
			$\sin^2\theta_{12}$&\quad\hskip -0.4 cm $0.321^{+0.018}_{-0.016}$\ , \quad $0.273\sim 0.379$ 
			& \quad\hskip -0.2 cm$0.321^{+0.018}_{-0.016}$ , \quad\  $0.273\sim 0.379$ \\
			\hline
			$\sin^2\theta_{13}$ $(10^{-2})$& \ \ \hskip -0.7 cm $2.155^{+0.090}_{-0.075}$ , \quad $1.89\sim 2.39$ &
			\quad\hskip -0.6 cm$2.140^{+0.082}_{-0.085}$ , \quad\ $1.89\sim 2.39$ \\
			\hline
		\end{tabular}
		\caption{ $1 \sigma$ and $3\sigma$ ranges of the global analysis of the neutrino oscillation experimental data
			for NH and IH, respectively \cite{deSalas:2017kay}. }
		\label{tab}
	\end{center}
\end{table}

In the next subsections,
we show the numerical results in the case of $\rm TM_1$ and $\rm TM_2$ for both NH and IH.

\subsection{Three cases of $\rm TM_1$ (2-3 family mixing) in NH}\label{23NHmodel}
\begin{figure}[h!]
	\begin{tabular}{ll}
		\begin{minipage}{0.5\hsize}
			\begin{center}
				\includegraphics[width=\linewidth]{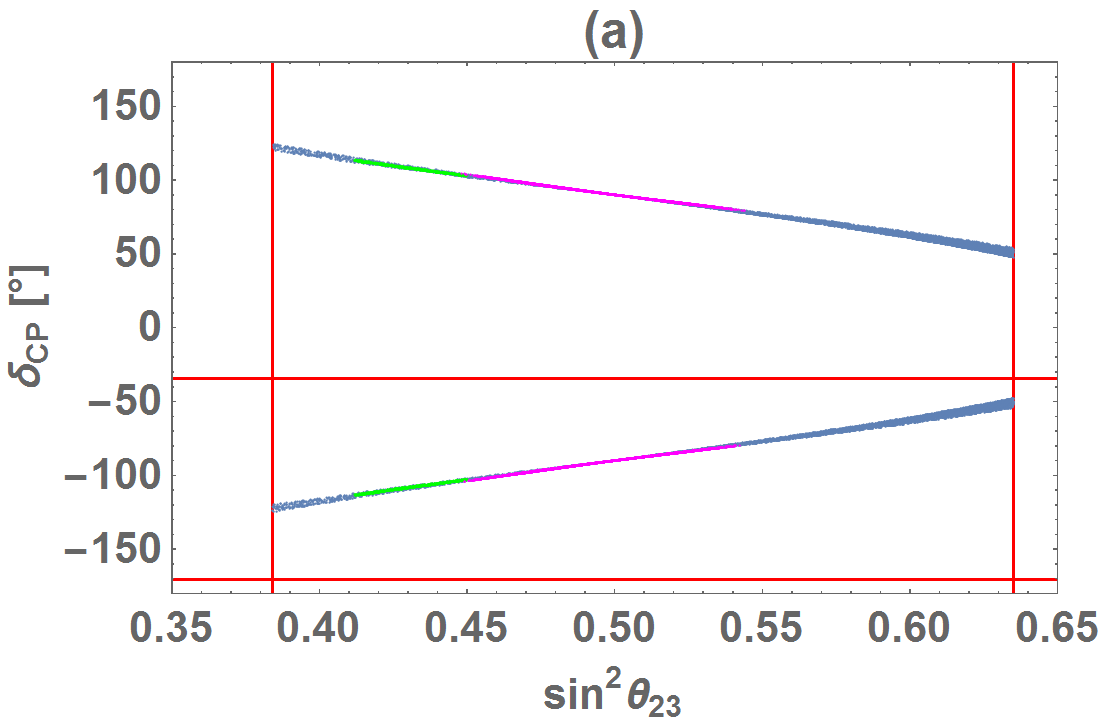}
			\end{center}
		\end{minipage}
		\begin{minipage}{0.5\hsize}
			\begin{center}
				\includegraphics[width=\linewidth]{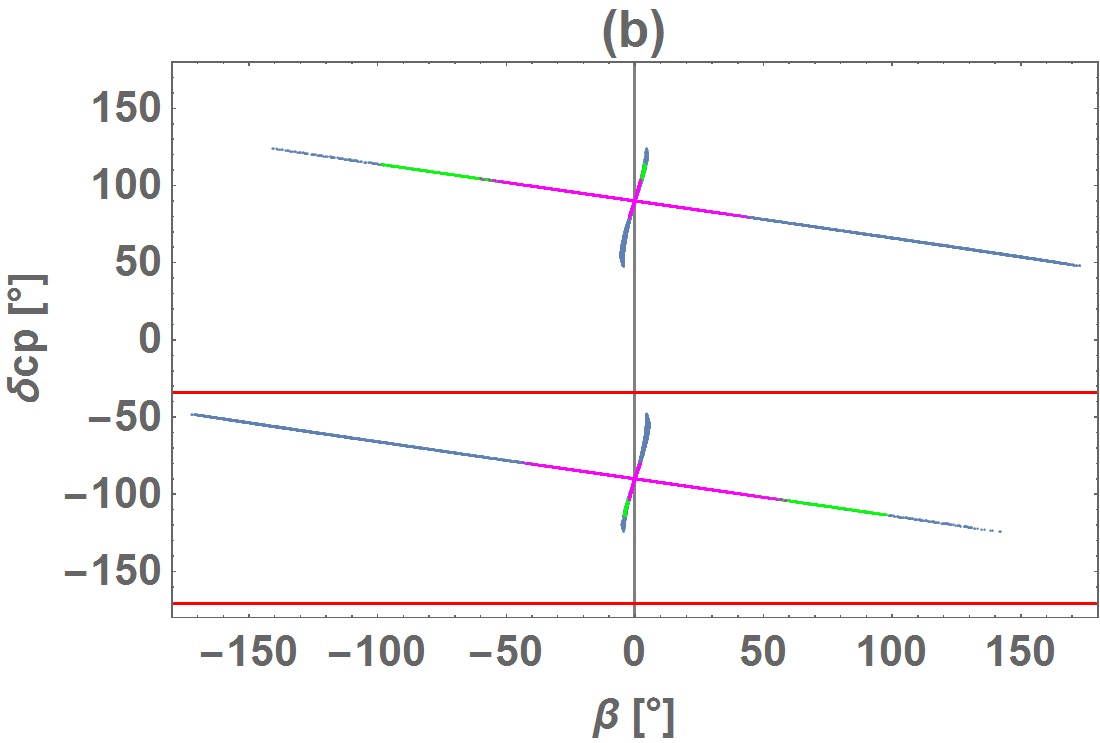}
			\end{center}
		\end{minipage}
	\end{tabular}
	\begin{tabular}{ll}
		\begin{minipage}{0.5\hsize}
			\begin{center}
				\includegraphics[width=\linewidth]{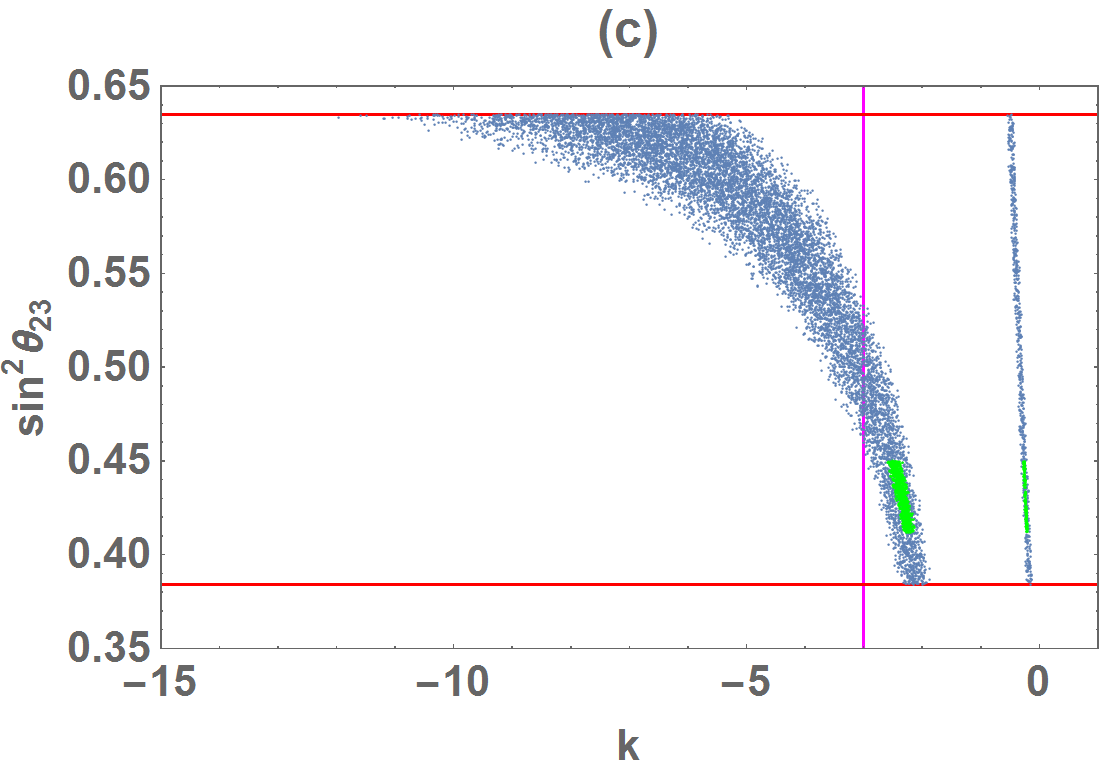}
			\end{center}
		\end{minipage}
		\begin{minipage}{0.5\hsize}
			\begin{center}
				\includegraphics[width=\linewidth]{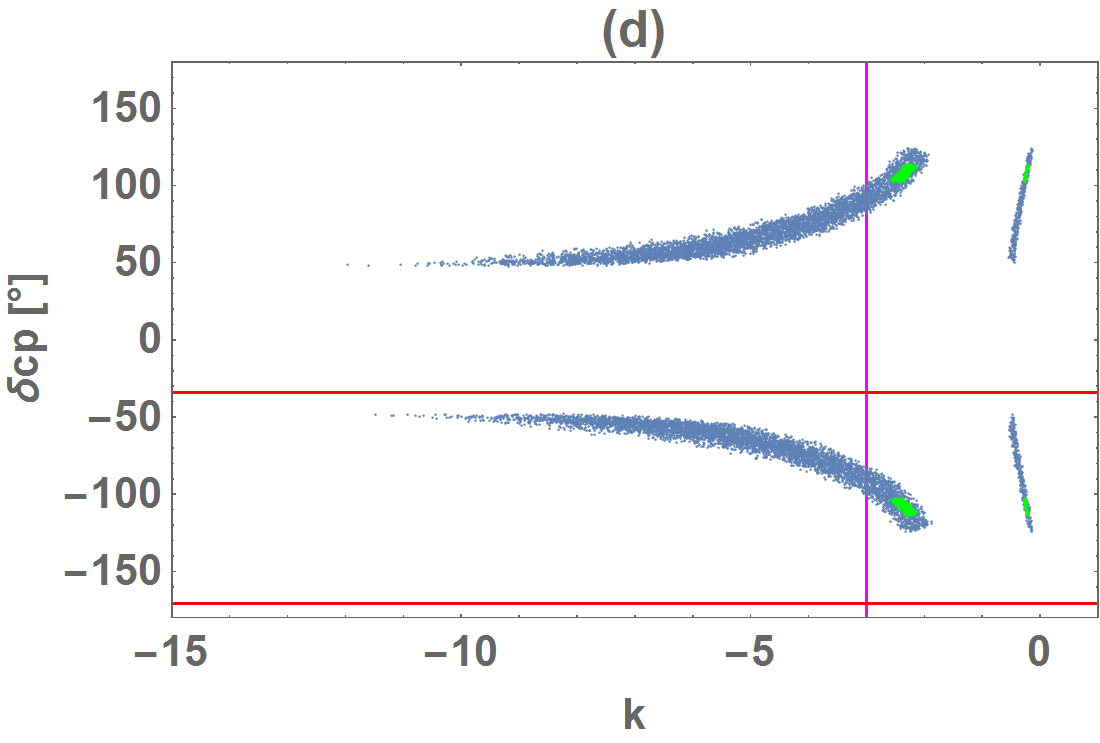}
			\end{center}
		\end{minipage}
	\end{tabular}
	\begin{tabular}{ll}
		\begin{minipage}{0.5\hsize}
			\begin{center}
				\includegraphics[width=\linewidth]{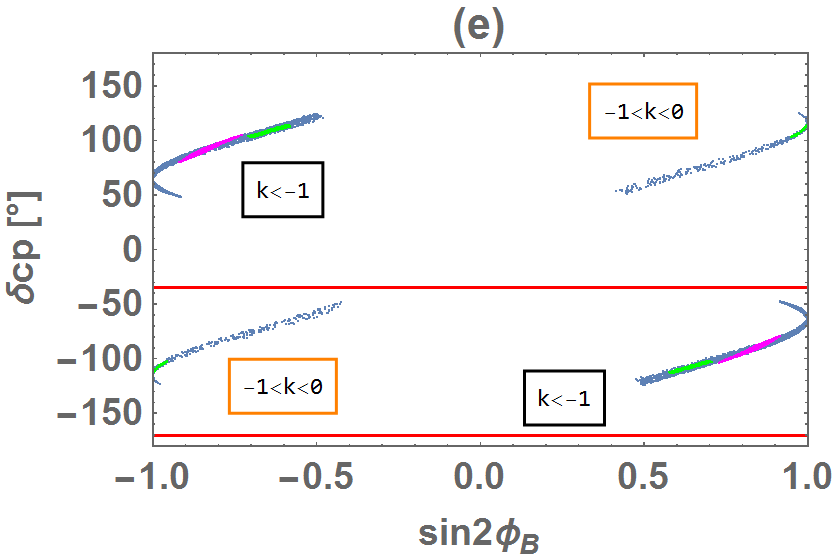}
			\end{center}
		\end{minipage}
		\begin{minipage}{0.5\hsize}
			\begin{center}
				\includegraphics[width=\linewidth]{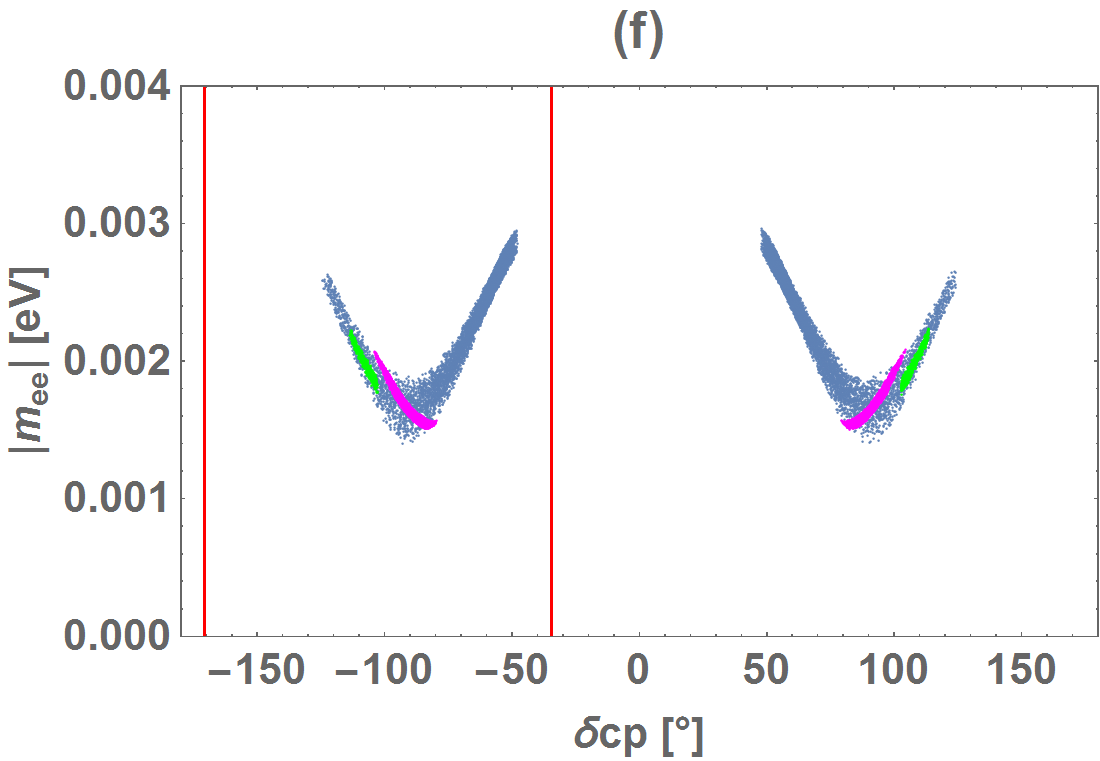}
			\end{center}
		\end{minipage}
	\end{tabular}
	\caption{
	Blue (green) dots denote predictions for case I within $3\sigma \ (1\sigma)$ range. 
	The red lines for $\sin^2\theta_{23}$ and $\delta_{CP}$ denote
	the  experimental bounds of $3\sigma$ (global analyses) and $2\sigma$ (T2K) ranges, respectively:
		(a)  $\delta_{CP}$ versus $\sin^2\theta_{23}$,
		(b)  $\delta_{CP}$ versus  $\beta$,
		(c) $\sin^2\theta_{23}$ versus $k$,
		(d) $\delta_{CP}$ versus $k$,
		(e)  $\delta_{CP}$ versus $\sin 2\phi_B$,
		and (f)  $|m_{ee}|$ versus  $\delta_{CP}$.
		The magenta dots in (a), (b), (e) and (f) denote the predictions of the Littlest seesaw model.
		The vertical magenta lines of $k=-3$ in (c) and (d) also denote its predictions.
	}
	\label{fig:23NHcase1}
\end{figure}

Let us show numerical results of case I, II, and III in subsection 2.1.
The most interesting case is the case I.
The case I corresponds to one of the one-zero texture models,
where the $(1,1)$ element of the Dirac neutrino mass matrix is zero.
There are two  free parameters,  $k$ and  $\phi_B$  after inputting data of neutrino masses in Table 1.
As seen in Fig.~\ref{fig:23NHcase1}(a),
the prediction of $\delta_{CP}$ versus $\sin^2\theta_{23}$ is same as the previous result of $\rm TM_1$~\cite{Shimizu:2014ria},
where the neutrino mass matrix and the neutrino mass hierarchy have not been specified.
Unless the magnitude of the dimensionless parameter $k$ is fixed,
the predicted $\sin^2\theta_{23}$ is allowed to take all values within the $3\sigma$ range of the experimental data
while $\delta_{CP}$ is in the region of $\pm(45^\circ \sim 125^\circ )$
 depending on $\sin^2\theta_{23}$ for the inputting data within $3\sigma$.
 It is helpful to show the predictions for  the inputting data within $1\sigma$.
The predicted $\delta_{CP}$ is in the region of $\pm(100^\circ\sim 115^\circ )$  for the  $1\sigma$ range of  $\sin^2\theta_{23}=0.412-0.450$.
It is remarked that $|\delta_{CP}|$ should be larger than $90^\circ$
if the observed $\theta_{23}$ is in the first octant.
 This prediction is consistent with the recent data of T2K,
which present  the allowed $2\sigma $ range for the CP violating phase, $\delta_{CP}$, such as  $(-171^\circ, -34.4^\circ)$ for NH \cite{T2K}.

As well known, the predicted $\sin^2 \theta_{12}$ is $0.317\sim 0.320$ \cite{Shimizu:2014ria},
which is a characteristic one for $\rm TM_1$ independent of details of the neutrino mass matrix.
This predicted value is inside the region of the experimental data of $1\sigma$ range.

We also show the result for the Majorana phase $\beta$ in Fig.~\ref{fig:23NHcase1}(b). 
For  the inputting data within $3\sigma$,  $\beta=0$ is allowed,
but $\beta=0$ is excluded for  the inputting data with $1\sigma$.
It is found  that $\beta$ is zero when the Dirac CP violating phase $\delta_{CP}$ is $\pm \pi/2$.
The remaining Majorana phase $\alpha$ is arbitrary because of $m_1=0$.

It is remarked that our predictions depend on the parameter $k$.
As seen in Fig.~\ref{fig:23NHcase1}(c),
it is allowed in two separated region, 
$k\simeq -12\sim -2$ and $k\simeq -0.1\sim -0.5$
for  the inputting data within $3\sigma$.
These regions are related to inverse values each other.
The predicted value of $\sin^2\theta_{23}$ crucially depends on $k$ as seen in Fig.~\ref{fig:23NHcase1}(c).
Indeed, for the inputting data within $1\sigma$, $k$ is severely constrained
around $-2.5$ and $-0.4$.

We also present $k$ dependence of $\delta_{CP}$ in Fig.~\ref{fig:23NHcase1}(d).
If the observed $|\delta_{CP}|$ is larger than $90^\circ$,
$k$ is larger than $-3$.
If $\delta_{CP}$ is observed accurately as well as $\sin^2\theta_{23}$, $k$ is determined with two-fold ambiguity.
The $k$ dependence is important  because textures of neutrino mass matrix
 are given by fixing  $k$ as discussing later.

We present the $\sin 2\phi_B$ dependence of $\delta_{CP}$ in
 Fig.~\ref{fig:23NHcase1}(e), where the prediction is  classified by $k$.
 It is remarked that  the sign of $\delta_{CP}$
 depends on both $\sin 2\phi_B$ and $k$.
 The sign of $\delta_{CP}$ is determined by the sign of $(\sin 2\phi_B)$
  for $k>-1$ while by the sign of  $(-\sin 2\phi_B)$ for $k<-1$.
 This result is important for model-buildings.

We  show the effective mass for the $0\nu\beta\beta$ decay $|m_{ee}|$ versus $\delta_{CP}$ in Fig.~\ref{fig:23NHcase1}(f),
where $|m_{ee}|$ is around $1.5\sim 3.0$ meV.
It is  minimal  if the CP violating phase  $\delta_{CP}$ is  $\pm\pi/2$.
The correlation between $|m_{ee}|$ and $\delta_{CP}$ is helpful to test 
our model for the Dirac neutrino mass matrix.

Finally, we discuss the Littlest seesaw model~\cite{King:2015dvf}-\cite{King:2016yvg}
 in Eq.~(\ref{King}) of subsection 2.1.
This corresponds to $k=-3$ in the case I.
As already presented in Ref.~\cite{King:2015dvf},
this model leads to the predictions around $\sin^2\theta_{23}=1/2$ and the maximal $\delta_{CP}=\pm \pi/2$.
We show the results of the Littlest seesaw model by the color of magenta in Figs.~\ref{fig:23NHcase1}(a)-(f).
As seen in Fig.~\ref{fig:23NHcase1}(a),
the predicted $\sin^2\theta_{23}$ and $\delta_{CP}$ are $0.451\sim 0.544 $ and $\pm(80^\circ \sim 105^\circ )$, respectively.
As seen in Fig.~\ref{fig:23NHcase1}(b),
the Majorana phase $\beta$ is in the range from $-60^\circ$ to $60^\circ$.
In both Figs.~\ref{fig:23NHcase1}(c) and (d),
the vertical magenta lines of $k=-3$ denote the predictions of the Littlest seesaw model.
The effective mass $|m_{ee}|$ is in the minimal regions in Fig.~\ref{fig:23NHcase1}(f).

We propose new Littlest seesaw models by observing the result of  Fig.~\ref{fig:23NHcase1}.
For example, $k=-5$ or $-2$ $(-0.2 \ {\rm or}\  -0.5)$ give different predictions
for $\sin^2\theta_{23}$ and $\delta_{CP}$.
The Dirac neutrino mass matrices are
\begin{equation}
M_D=
\begin{pmatrix}
0 & 2f \\
b & 5f \\
-b & -f
\end{pmatrix}, \qquad
\begin{pmatrix}
0 & 2f \\
b & -f \\
-b & 5f
\end{pmatrix}, \qquad
\begin{pmatrix}
0 & f \\
b & 4f \\
-b & -2f
\end{pmatrix}, \qquad
\begin{pmatrix}
0 & f \\
b & -2f \\
-b & 4f
\end{pmatrix},
\label{others}
\end{equation}
which will be testable in the future experiments.
Thus, our analyses present new Dirac neutrino mass matrices,
which suggest new models of the lepton flavors.

\begin{figure}[h!]
	\begin{tabular}{cc}
		\begin{minipage}{0.5\hsize}
			\begin{center}
				\includegraphics[width=\linewidth]{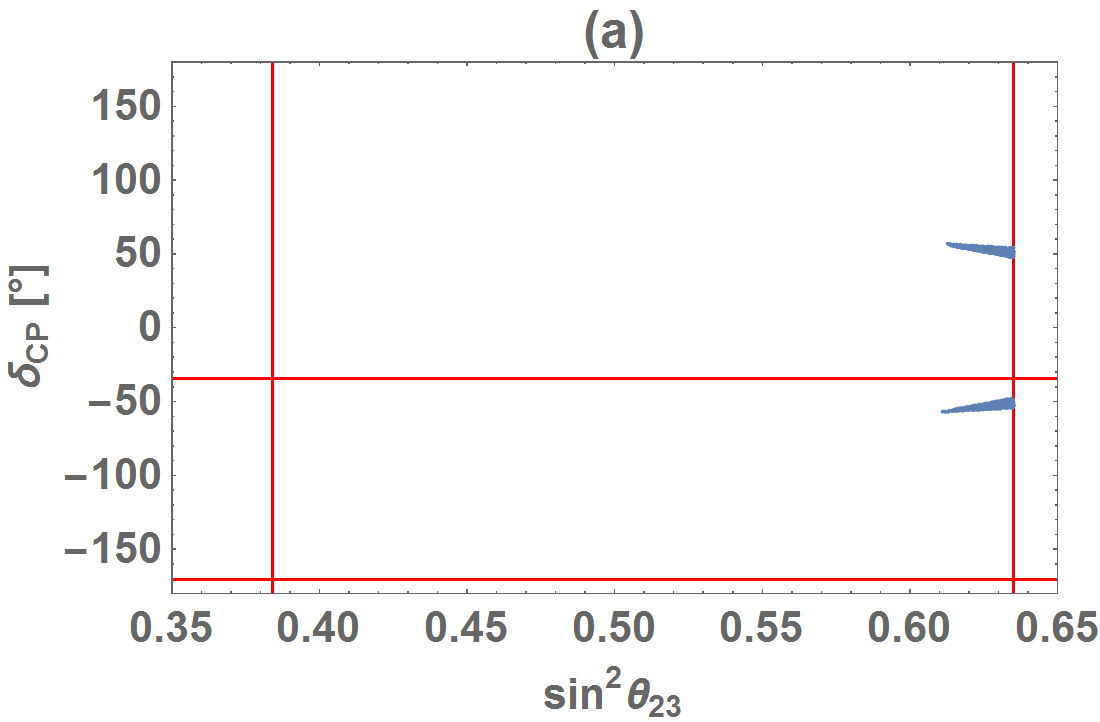}
			\end{center}
		\end{minipage}
		\begin{minipage}{0.5\hsize}
			\begin{center}
				\includegraphics[width=\linewidth]{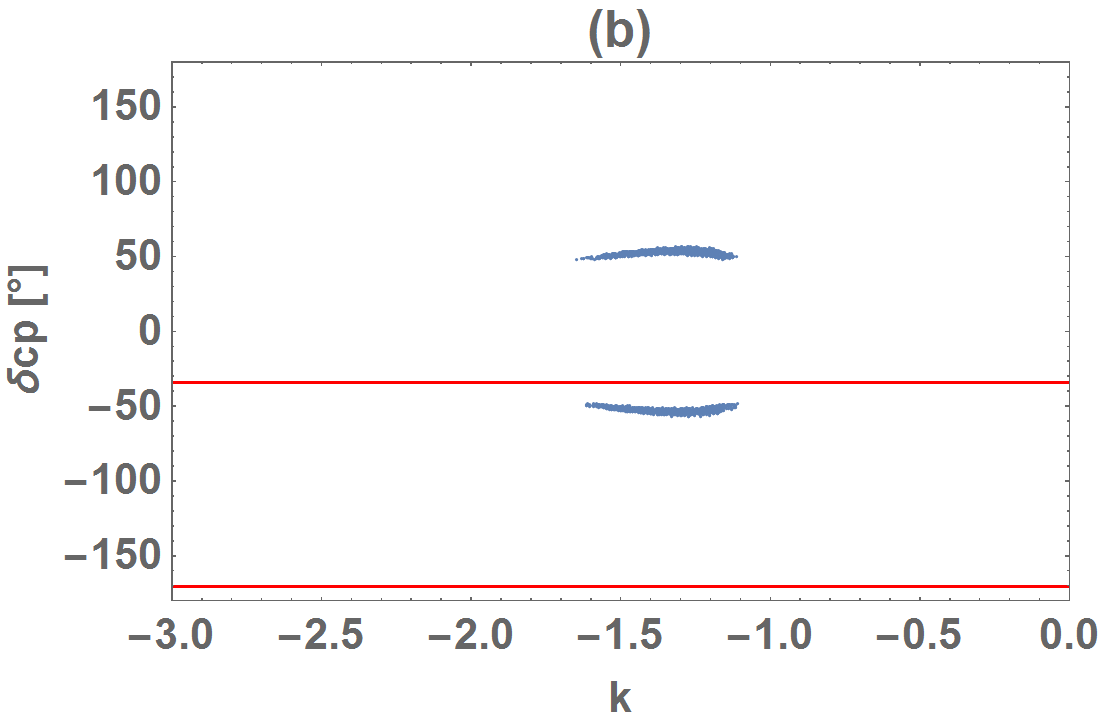}
			\end{center}
		\end{minipage}
	\end{tabular}
	\caption{
	Predictions for case II, where the red lines for $\sin^2\theta_{23}$ and $\delta_{CP}$ denote the experimental bounds of $3\sigma$ (global analyses) and $2\sigma$ (T2K) ranges, respectively:
	 (a)  $\delta_{CP}$ versus $\sin^2\theta_{23}$ and
	 (b) $\delta_{CP}$ versus $k$.}
	\label{fig:23NHcase2}
\end{figure}
\begin{figure}[h!]
	\begin{tabular}{cc}
		\begin{minipage}{0.5\hsize}
			\begin{center}
				\includegraphics[width=\linewidth]{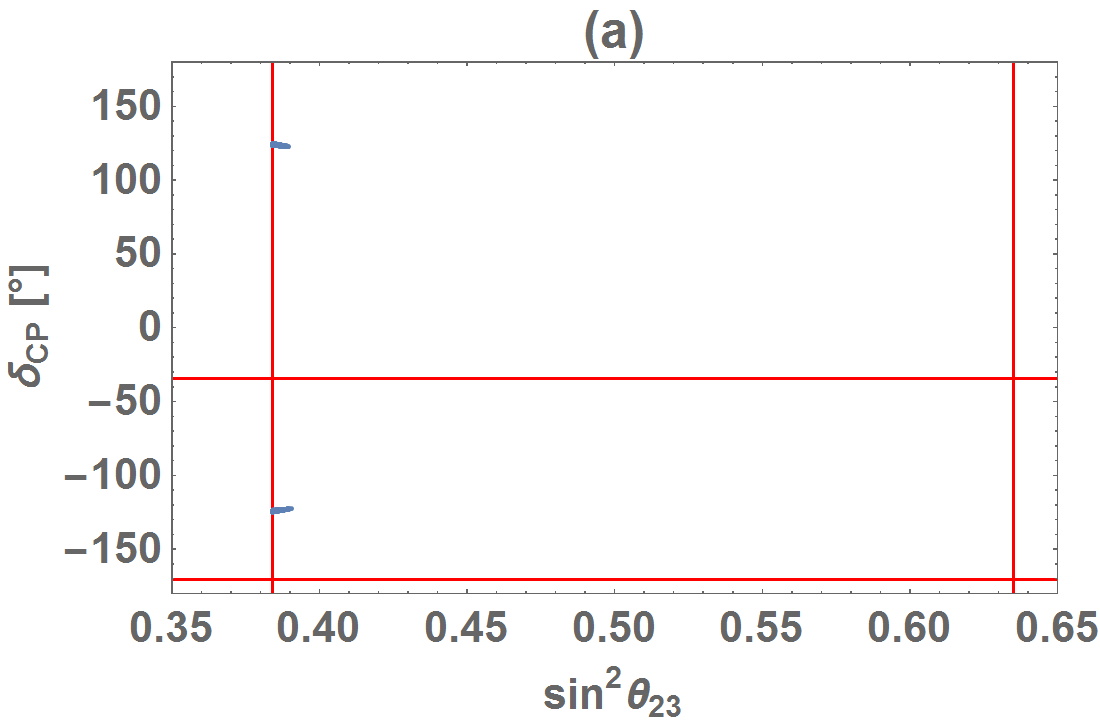}
			\end{center}
		\end{minipage}
		\begin{minipage}{0.5\hsize}
			\begin{center}
				\includegraphics[width=\linewidth]{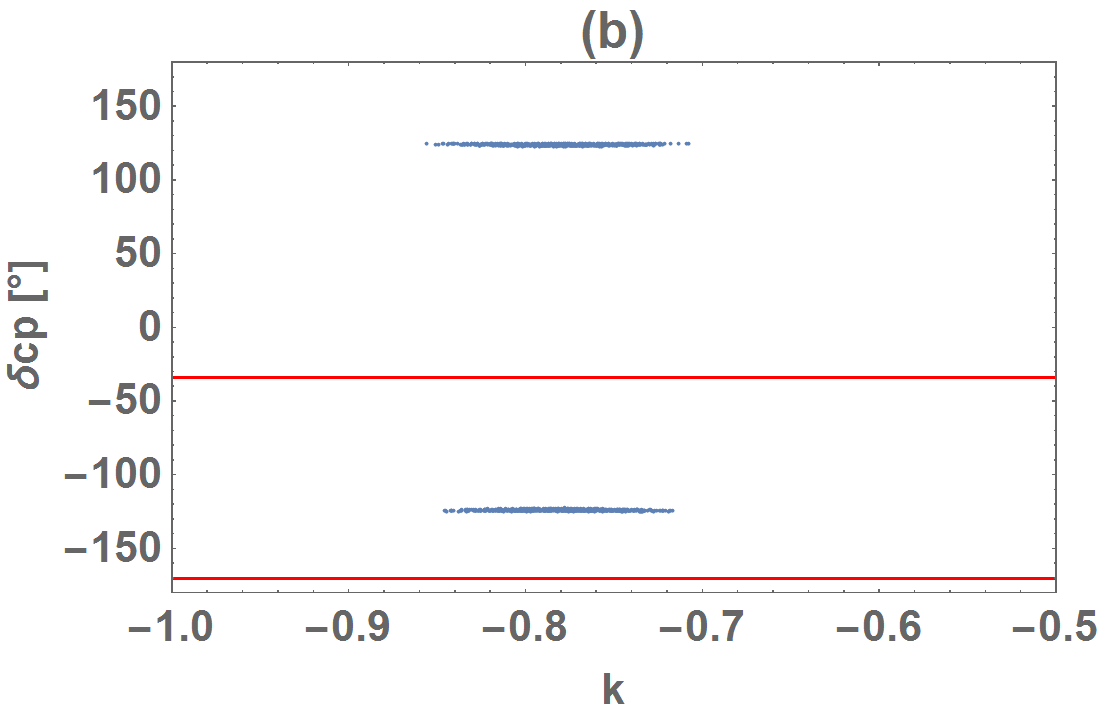}
			\end{center}
		\end{minipage}
	\end{tabular}
	\caption{
	Predictions for case III, where the red lines for $\sin^2\theta_{23}$ and $\delta_{CP}$ denote the  experimental bounds of $3\sigma$ (global analyses) and $2\sigma$ (T2K) ranges, respectively:
		(a)  $\delta_{CP}$ versus $\sin^2\theta_{23}$ and
		(b) $\delta_{CP}$ versus $k$.}
	\label{fig:23NHcase3}
\end{figure}
\begin{figure}[h!]
	\begin{center}
		\includegraphics[width=0.55\linewidth]{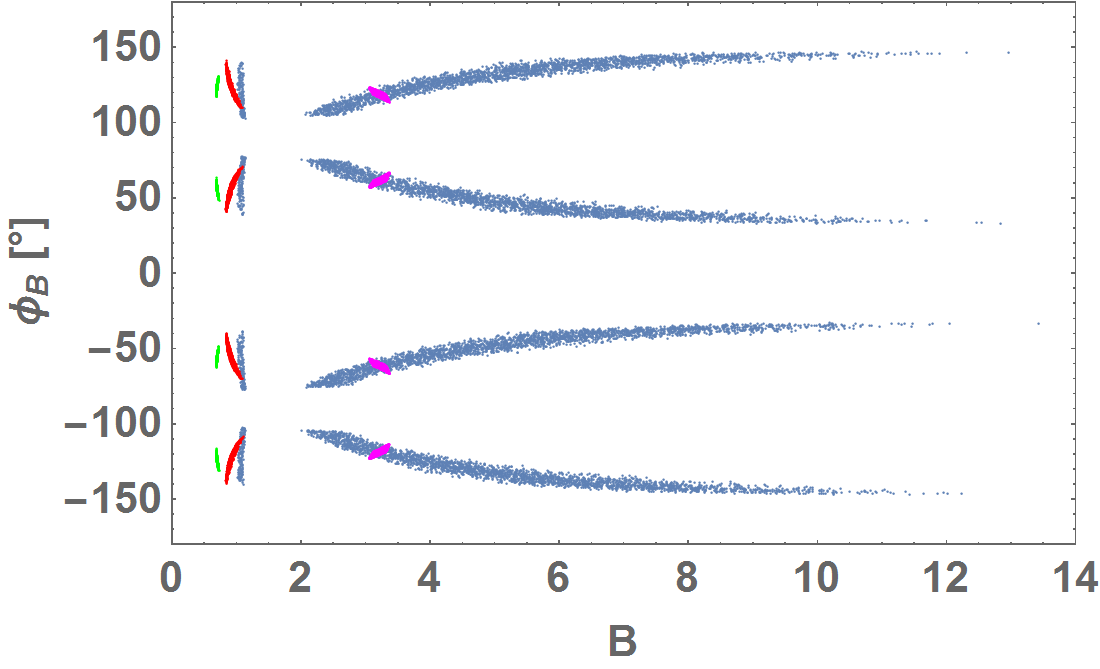}
	\end{center}
\caption{
Allowed regions in $B$-$\phi_B$ plane for $\rm TM_1$ in  NH.
The blue, red, green, and magenta dots denote the case I, II, III, and the Littlest seesaw model, respectively.}
\label{fig:BphiB}
\end{figure}
\begin{figure}[h!]
	\begin{tabular}{cc}
		\begin{minipage}{0.5\hsize}
			\begin{center}
				\includegraphics[width=\linewidth]{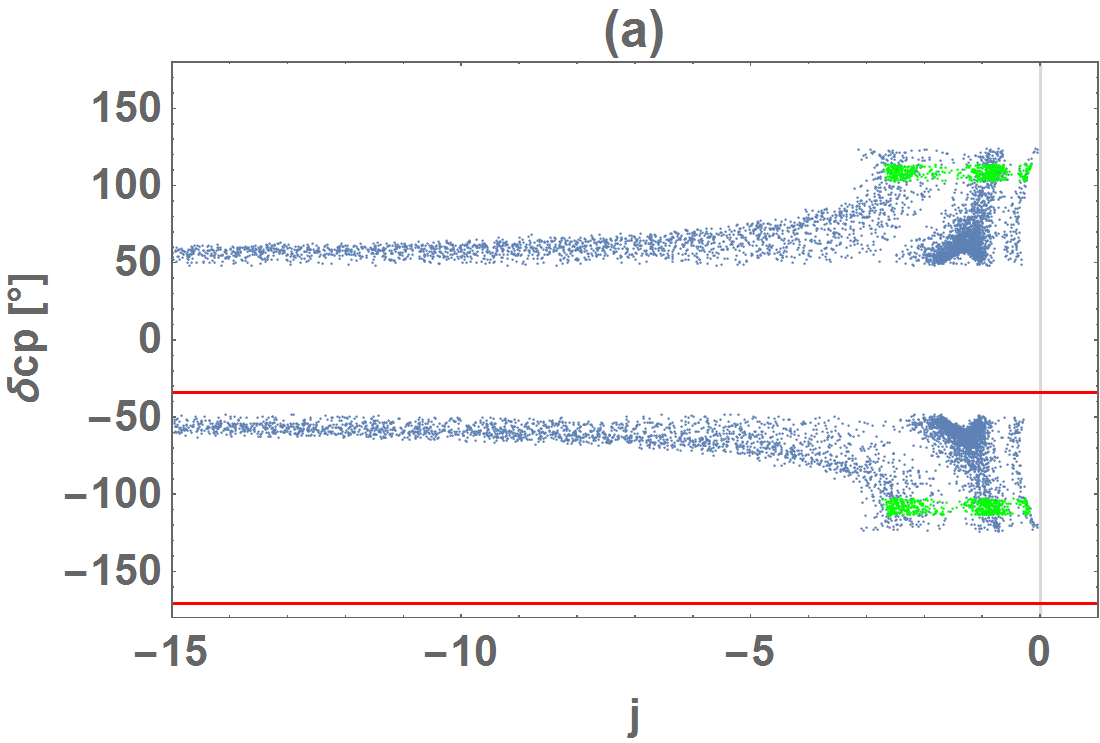}
			\end{center}
		\end{minipage}
		\begin{minipage}{0.5\hsize}
			\begin{center}
				\includegraphics[width=\linewidth]{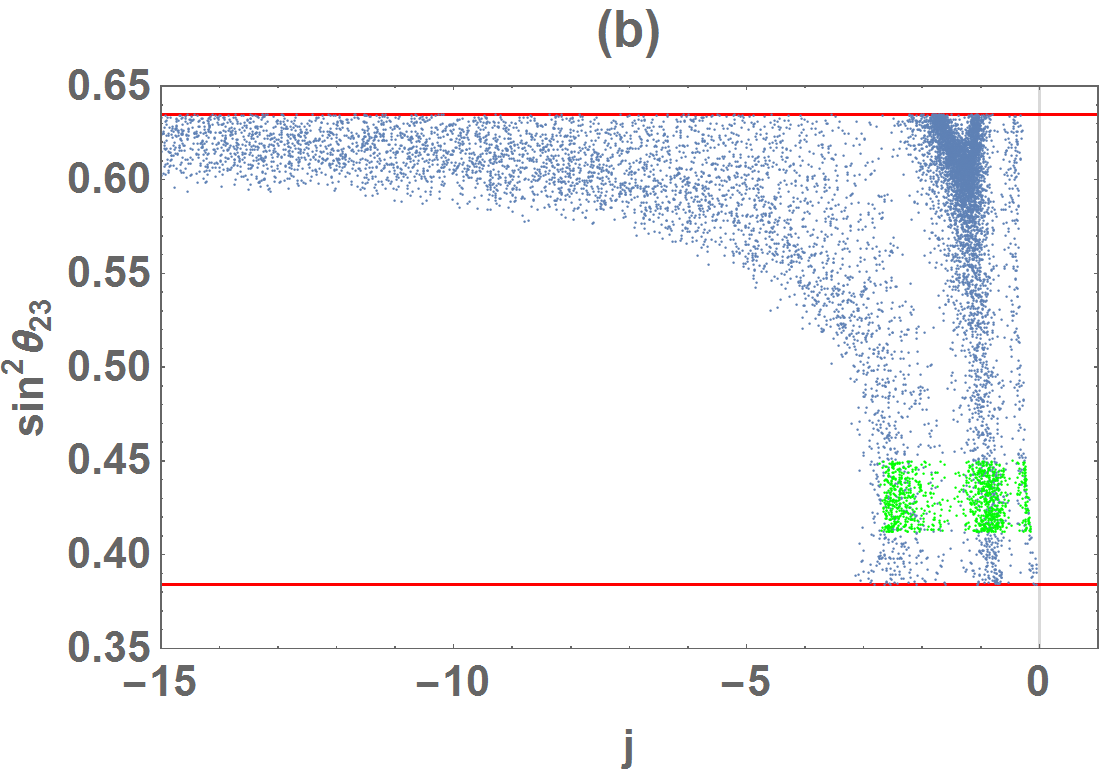}
			\end{center}
		\end{minipage}
	\end{tabular}
	\begin{tabular}{ll}
		\begin{minipage}{0.5\hsize}
			\begin{center}
				\includegraphics[width=\linewidth]{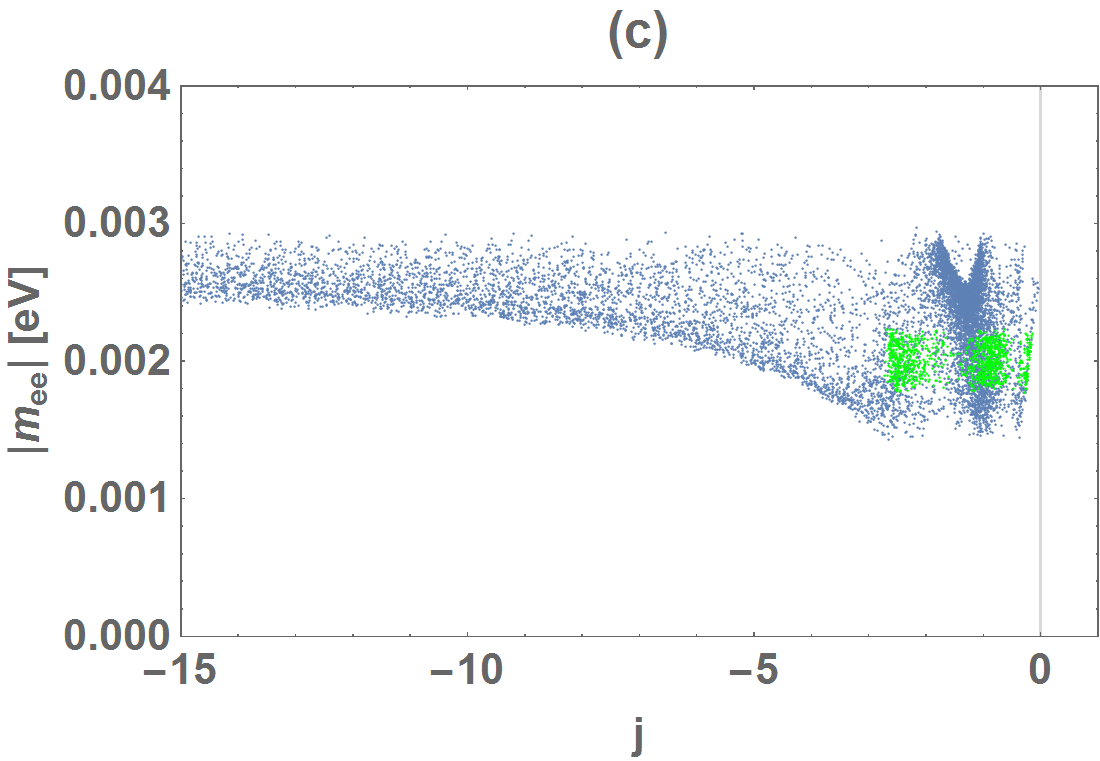}
			\end{center}
		\end{minipage}
		\begin{minipage}{0.5\hsize}
			\begin{center}
				\includegraphics[width=\linewidth]{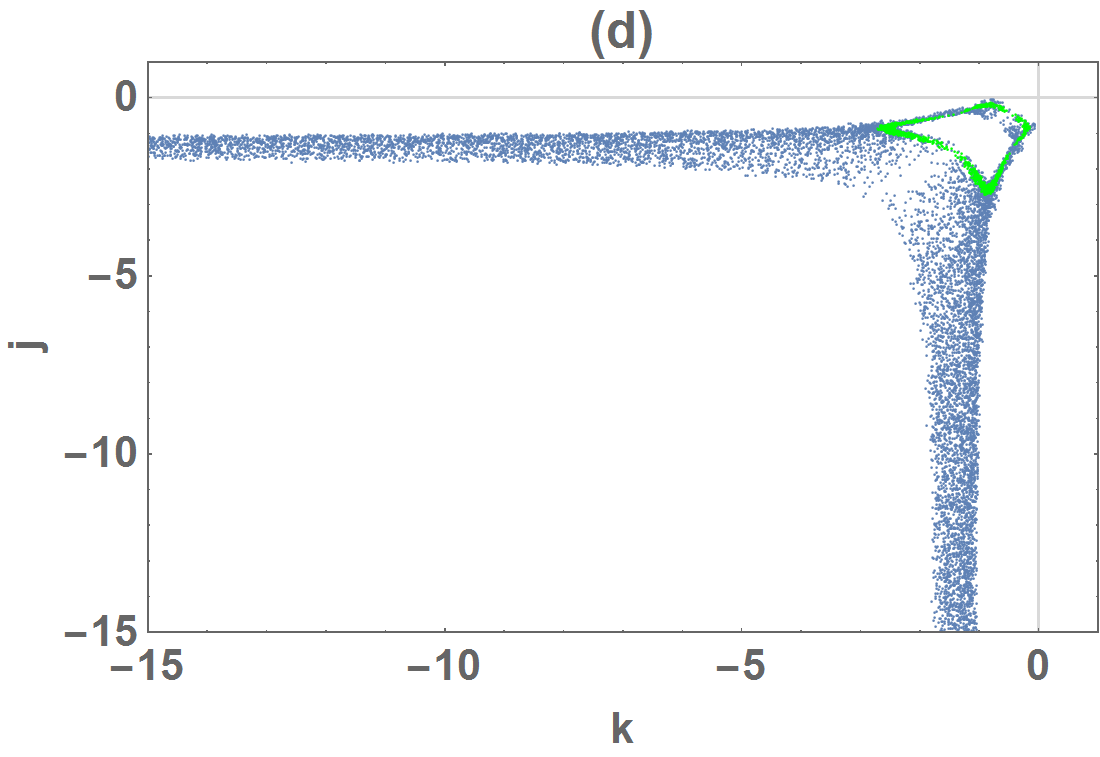}
			\end{center}
		\end{minipage}
	\end{tabular}
	\caption{Blue (green) dots denote predictions for  arbitrary  $j$ 
		in  $\rm TM_1$(NH)  within $3\sigma \ (1\sigma)$. The 
		red lines for $\sin^2\theta_{23}$ and $\delta_{CP}$ denote the  experimental bounds of $3\sigma$ (global analyses) and $2\sigma$ (T2K) ranges, respectively:
		(a)  $\delta_{CP}$ versus $j$,
		(b) $\sin^2\theta_{23}$ versus $j$,
		(c) $|m_{ee}|$ versus  $j$,
		and (d) $k$ versus $j$.
	}
	\label{fig:23NHgeneral}
\end{figure}


Next, we discuss the numerical results for the case II, where
 the $(2,1)$ element of the Dirac neutrino mass matrix is zero.
In Fig.~\ref{fig:23NHcase2}(a),
the predicted $\sin^2\theta_{23}$ is restricted near the upper bound of the $3\sigma$ range of the experimental data
and  the CP violating phase is $\delta_{CP}\simeq\pm 50^\circ$.
The case II may be excluded if the more precise data of $\sin^2\theta_{23}$ are available in the near future.

Although the prediction of  $\delta_{CP}$ in Fig.~\ref{fig:23NHcase2}(a) seems to be
in the partial regions of the one in the case I,
the  parameter $k$ of the case II is much different from the range of  $k$ in case I, that is, $k=-1.65\sim -1.10$ as seen in Fig.~\ref{fig:23NHcase2}(b).
We find  that the sign of $\delta_{CP}$ is determined 
 by the sign of $(\sin 2\phi_B)$ in the region of  $k=-1.65\sim -1.10$.

We note the Majorana phase and the effective mass for the $0\nu\beta\beta$ decay.
The Majorana phase $\beta$ is restricted to the region around $\pm 5^\circ$ and $\pm (140^\circ \sim 170^\circ )$.
The effective mass $|m_{ee}|$ is predicted to be $2.6\sim 3.0$ meV.


In the case III, the $(3,1)$ element of the Dirac neutrino mass matrix is zero.
In contrast to the case II,
the mixing angle $\sin^2\theta_{23}$ is strongly restricted to the lower bound of the $3\sigma$ range as seen in Fig.~\ref{fig:23NHcase3}(a).
The case III is almost excluded in the present status of the neutrino oscillation experiments.
We find  $k=-0.86\sim -0.71$. In this region,
the Dirac CP violating phase $\delta _{CP}$ is predicted in the region around $\pm 125^\circ $.
The Majorana phase $\beta$ is predicted in the region around $\pm 5^\circ $ and $\pm 140^\circ $.
It is also remarked that the sign of $\delta_{CP}$ is determined 
by the sign of $(-\sin 2\phi_B)$ in the  region of $k=-0.86\sim -0.71$.
The predicted effective mass $|m_{ee}|$ is around $2.7$ meV.

Finally in this subsection,
we discuss allowed region of the parameters $B$ and $\phi_B$ in Eq.~(\ref{parameters}).
We show the combined results for cases I, II, III, and the Littlest seesaw model in Fig.~\ref{fig:BphiB},
where the blue, red, green, and magenta dots denote the case I, II, III, and the Littlest seesaw model, respectively.
Thus, the allowed regions of those parameters are clearly different each other.

\subsection{General case  for $\rm TM_1$ (2-3 family mixing) in NH }\label{23NHgeneralmodel}
In the previous subsection, $j\equiv b/c$ has been fixed.
The cases I, II, and III are specific cases for $j$,
which correspond to $j= -1$, $-\infty$, and $0$, respectively.
In order to show $j$ dependence of our predictions in detail,
we discuss the case in which $j$ is not specified as in Eq.(\ref{left-handed-1}).

We show the prediction of $\delta_{CP}$ versus $j$ in Fig.~\ref{fig:23NHgeneral}(a),
where $j$ is allowed in negative value.
As $|j|$ increases, $\delta_{CP}$ reaches $\pm 45^\circ$ asymptotically.
We show the prediction of $\sin^2\theta_{23}$ versus $j$ in Fig.~\ref{fig:23NHgeneral}(b).
As far as $j$ is larger than $-3$,
$\sin^2\theta_{23}$ is allowed to take all values within the $3\sigma$ range of the experimental data.
However, it is predicted to be larger than $0.5$ when $j$ is smaller than $-3$.
It is also easily seen that
the cases II $(j=-\infty)$ and III $(j=0)$ are marginal for the prediction of $\sin^2\theta_{23}$ as seen in Fig.~\ref{fig:23NHgeneral}(b).
We show the prediction of $|m_{ee}|$ versus $j$ in Fig.~\ref{fig:23NHgeneral}(c).
We find that the predicted region of $|m_{ee}|$ never enlarge
compared with those predictions of cases I, II, and III even if $j$ is taken to be arbitrary.
Finally, we show the allowed region in $k$ and $j$ plane in Fig.~\ref{fig:23NHgeneral}(d).
It is verified numerically that $j$ and $k$ are symmetric for their exchange.
 The allowed regions within $1\sigma$ are rather narrow as seen in Fig.~\ref{fig:23NHgeneral}.
If the error-bar of the inputting data are reduced in the future, both $k$ and $j$
are expected to be constrained considerably.

 In these results, we take $j=b/c$ to be real although 
  the relative phase $b/c$ does not vanish in general as discussed below Eq.~(\ref{left-handed-1}).
We have checked numerically  that  an extra phase of  $b/c$
does not contribute the results in Fig.\ref{fig:23NHgeneral}  while it affects the $\phi_B$ dependences of
 $\delta_{CP}$ and $|m_{ee}|$ drastically. This situation is easily understood
  because the experimental data of mixing angles and mass square differences
  are input.
  
\subsection{$\rm TM_1$: 2-3 family mixing in IH}\label{23IHmodel}
Although the case of the NH is rather favored in the experiments of T2K and NO$\nu$A~\cite{Abe:2017vif}-\cite{Adamson:2017gxd}, 
the inverted hierarchy of neutrino masses is still allowed by the experimental data of the neutrino oscillations.
Let  us  show the numerical result of the IH case $(m_3=0)$ for the 2-3 family mixing in Eq.~(\ref{Mnu23-IH})
since the Dirac neutrino mass matrix is completely different from the one of NH.
In this case, the first column of the Dirac neutrino mass matrix is uniquely fixed as $(-2,1,1)^\mathrm{T}$.
There are two free parameters, $k$ and $\phi_k$, as seen in Eq.~(\ref{parametersIH}).
In contrast to the NH case, the ratio $e/f$ is not allowed to be real in order to obtain the CP violation.

\begin{figure}[h!]
	\begin{tabular}{cc}
		\begin{minipage}{0.5\hsize}
			\begin{center}
				\includegraphics[width=\linewidth]{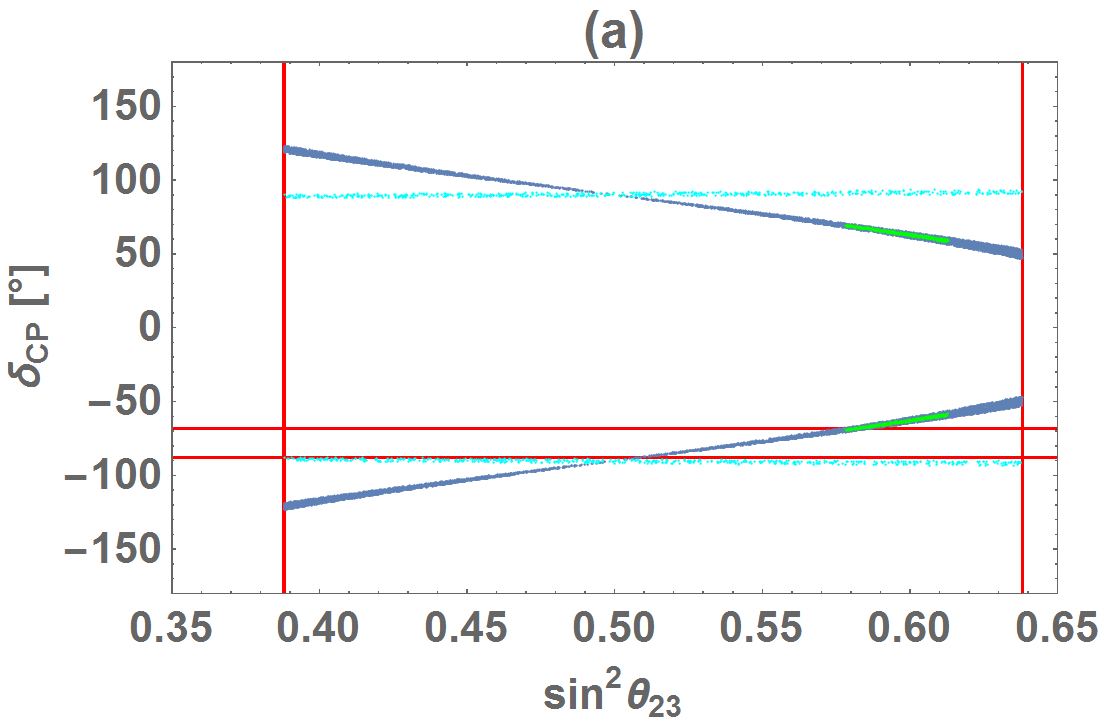}
			\end{center}
		\end{minipage}
		\begin{minipage}{0.5\hsize}
			\begin{center}
				\includegraphics[width=\linewidth]{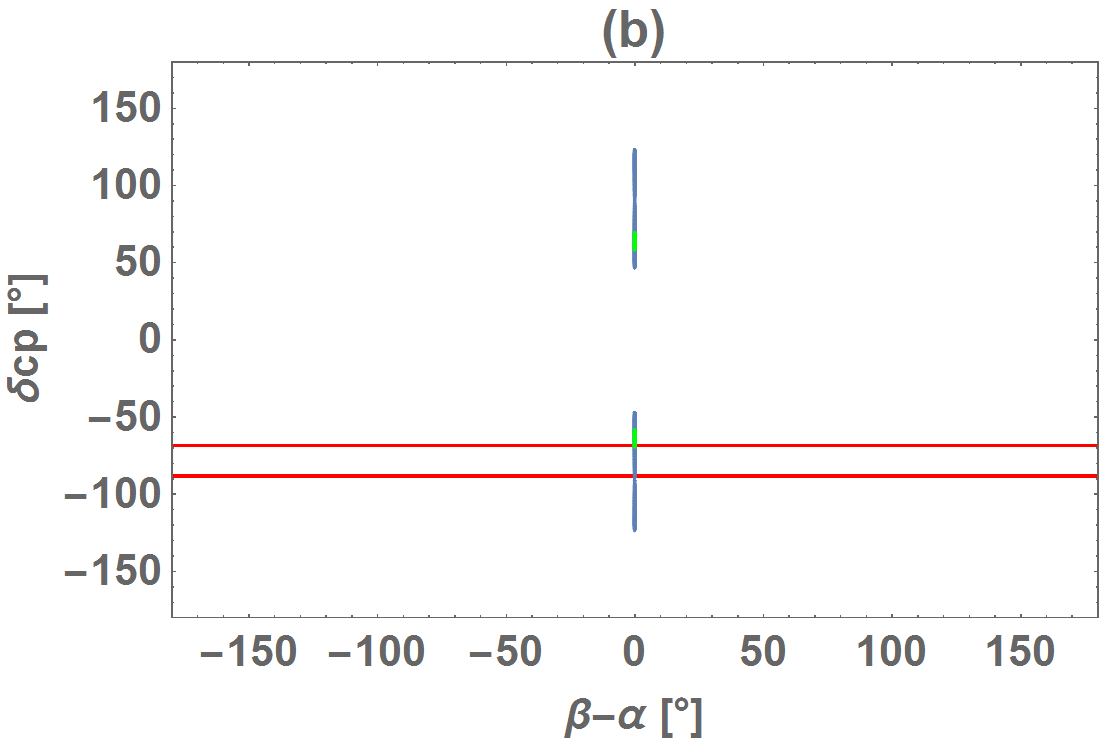}
			\end{center}
		\end{minipage}
	\end{tabular}
	\begin{tabular}{cc}
		\begin{minipage}{0.5\hsize}
			\begin{center}
				\includegraphics[width=\linewidth]{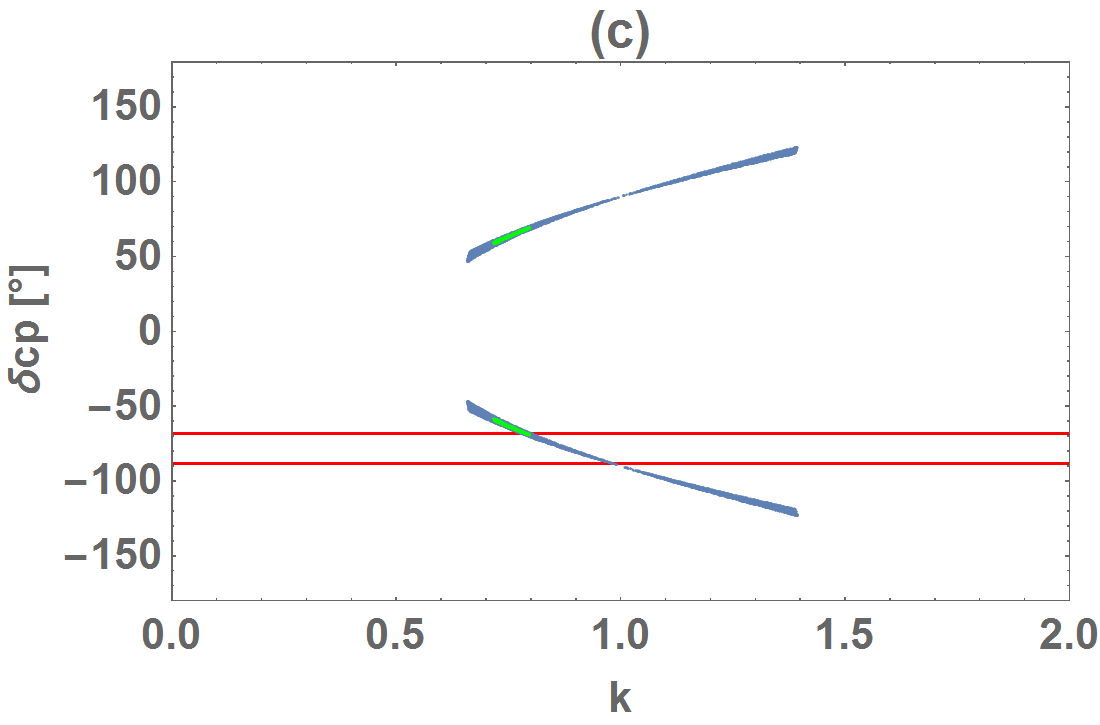}
			\end{center}
		\end{minipage}
		\begin{minipage}{0.5\hsize}
			\begin{center}
				\includegraphics[width=\linewidth]{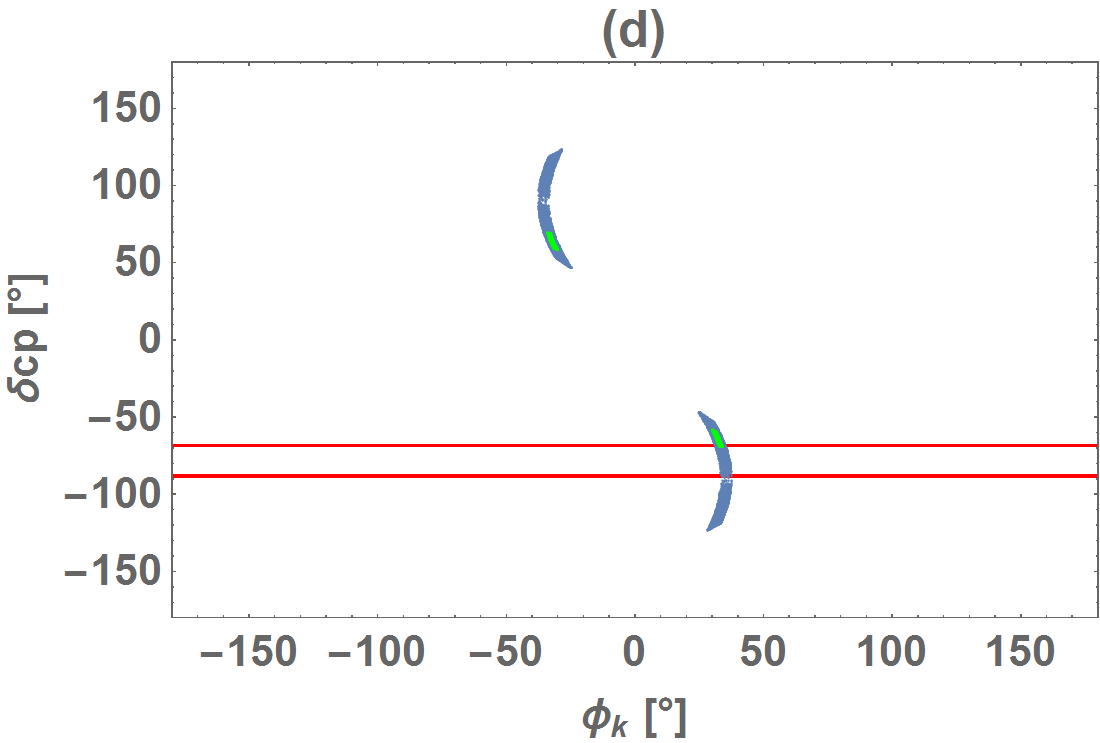}
			\end{center}
		\end{minipage}
	\end{tabular}
	\caption{ Blue (green) dots denote predictions for $\rm TM_1$ in IH
		with $3\sigma \ (1\sigma)$. The
		red lines for $\sin^2\theta_{23}$ and $\delta_{CP}$ denote the  experimental bounds of $3\sigma$ (global analyses) and $2\sigma$ (T2K) ranges, respectively:
		(a)  $\delta_{CP}$ versus $\sin^2\theta_{23}$,
		(b)  $\delta_{CP}$ versus difference of two Majorana phases $\beta-\alpha$,
		(c)  $\delta_{CP}$ versus $k$,
		and (d)   $\delta_{CP}$ versus $\phi_k$.
	The cyan dots in (a) denote the prediction of the two-zero texture \cite{Harigaya:2012bw}. }
	\label{fig:23IH}
\end{figure}

We show the predicted $\delta_{CP}$ versus $\sin^2\theta_{23}$ in Fig.~\ref{fig:23IH}(a).
The prediction is same as the case I of NH in Fig.~\ref{fig:23NHcase1}(a),
that is  $\pm(45^\circ \sim 125^\circ )$.
The recent T2K data present   $\delta_{CP}$  to be in the range of $(-88^\circ, -68^\circ)$ at $2\sigma$ range
for IH~\cite{T2K}.
We also show this range by red lines in the figures as an eye guide.
We plot $\delta_{CP}$ versus the difference of the two Majorana phases $\beta-\alpha$ in Fig.~\ref{fig:23IH}(b).
The difference of two Majorana phases $\beta-\alpha$ is almost zero.
That is $\alpha=\beta$.

We also show the $k$ and $\phi_k$ dependencies of $\delta_{CP}$ in Figs.~\ref{fig:23IH}(c) and (d).
The parameters are allowed in the narrow ranges $k=0.65\sim 1.40$ and
 $\phi_k=\pm (25^\circ \sim 38^\circ)$.
Thus, the structure of the Dirac neutrino mass matrix is restricted considerably.

The effective mass $|m_{ee}|$ is predicted to be  around $50$ meV
since $\beta-\alpha$ is almost zero and $|m_{ee}|$ does not depend on $\delta_{CP}$ because of $m_3=0$.

In the  context of the minimal seesaw model,
the two-zero texture for the Dirac neutrino mass matrix has been examined
in some works \cite{Frampton:2002qc,Harigaya:2012bw,Zhang:2015tea,Rink:2016vvl}.
 This texture is different from our ones.
However, this model is completely consistent with the experimental data of mixing angles and masses for IH \cite{Harigaya:2012bw}
although this texture is disfavored for NH without some corrections.
Since it is interesting to compare the predictions of two models,
 we add  its prediction of $\delta_{CP}$ 
in Figs.~\ref{fig:23IH}(a),
where $(1,1)$ and  $(2,2)$ elements of  the  $3\times 2$ 
 Dirac neutrino mass matrix are zero.
The predicted $\delta_{CP}$ is $\pm (88-93)^\circ$ in the whole range of 
 allowed $\sin^2\theta_{23}$.
 The case of vanishing  $(1,1)$ and  $(3,2)$ elements is also available,
  but the numerical prediction is almost unchanged.
 The prediction of the two-zero texture is a distinctive one from our ones.


\subsection{$\rm TM_2$: 1-3 family mixing in NH or IH}\label{13model}
In this subsection, we discuss the case of the 1-3 family mixing.
As well known, the predicted $\sin^2 \theta_{12}$ is close to $0.34$ \cite{Shimizu:2014ria},
which is a characteristic one for $\rm TM_2$.
The precise data of $\sin^2 \theta_{12}$ 
provides a crucial test for  $\rm TM_2$  as well as  $\rm TM_1$.

\begin{figure}[h!]
	\begin{tabular}{cc}
		\begin{minipage}{0.5\hsize}
			\begin{center}
				\includegraphics[width=\linewidth]{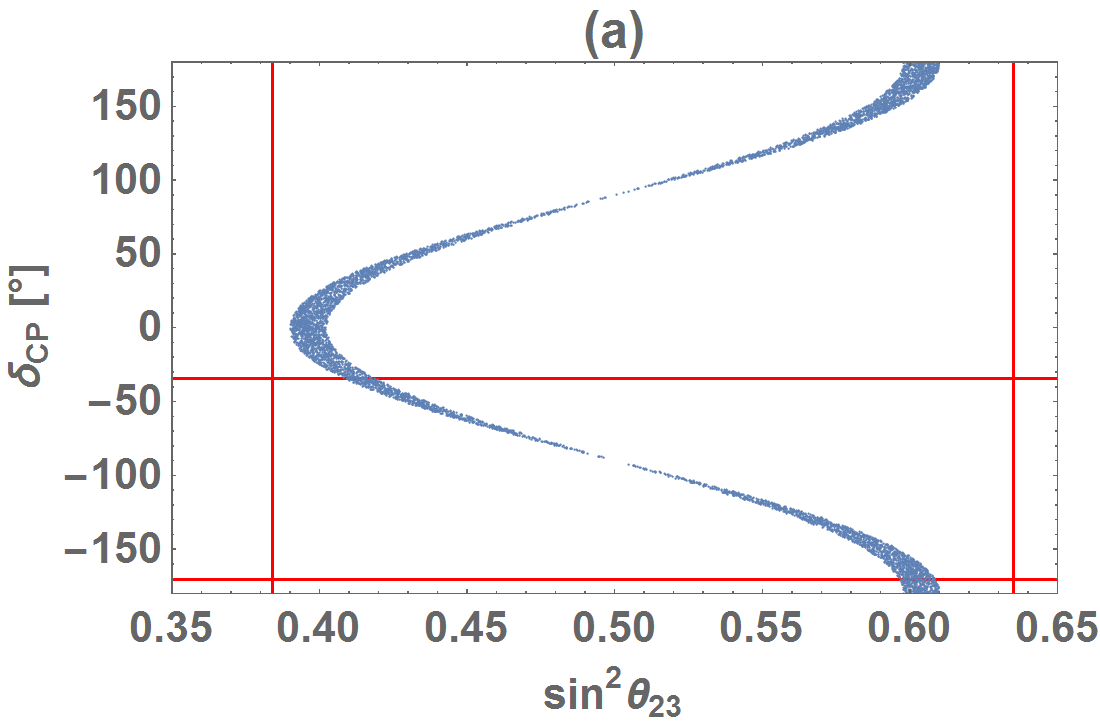}
			\end{center}
		\end{minipage}
		\begin{minipage}{0.5\hsize}
			\begin{center}
				\includegraphics[width=\linewidth]{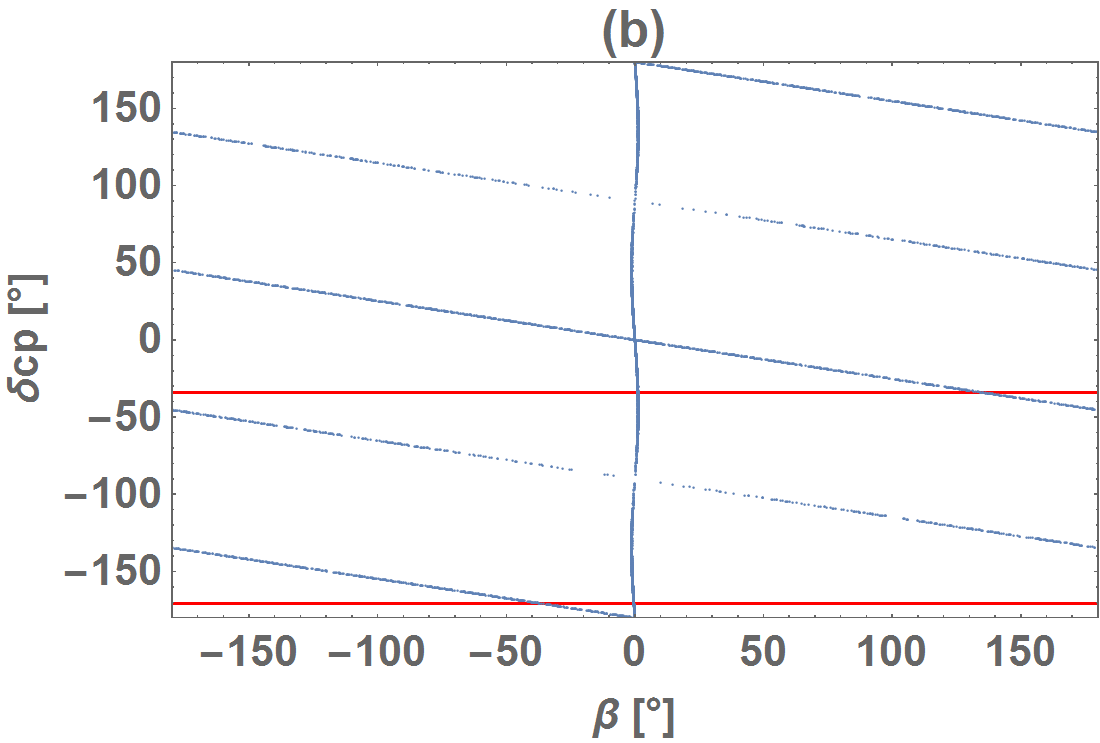}
			\end{center}
		\end{minipage}
	\end{tabular}
	\begin{tabular}{cc}
		\begin{minipage}{0.5\hsize}
			\begin{center}
				\includegraphics[width=\linewidth]{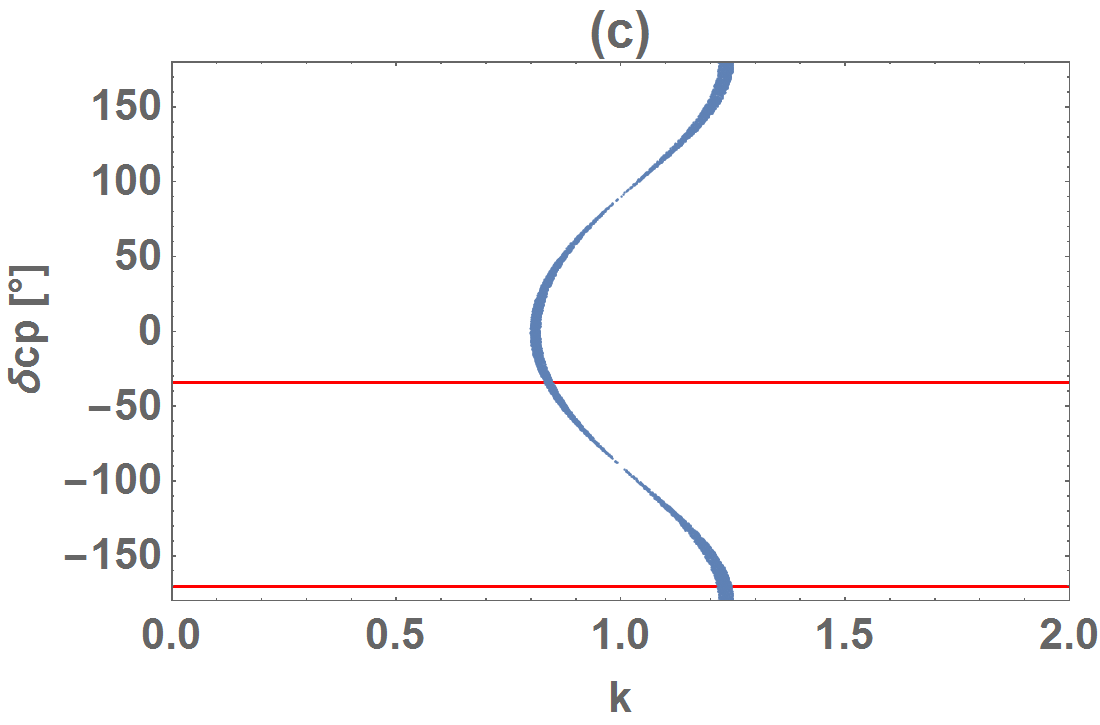}
			\end{center}
		\end{minipage}
		\begin{minipage}{0.5\hsize} 
				\begin{center}
				\includegraphics[width=\linewidth]{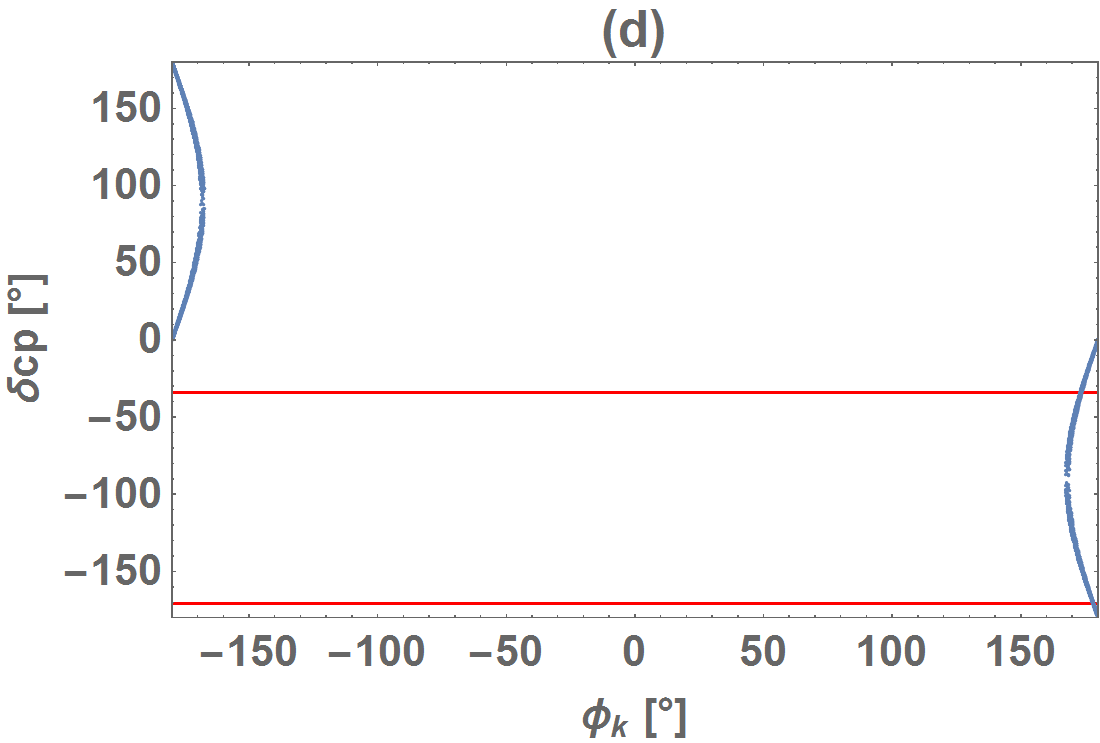}
		\end{center}
	\end{minipage}
	\end{tabular}
	\caption{
	Predictions for $\rm TM_2$ in  NH, where
the red lines for $\sin^2\theta_{23}$ and $\delta_{CP}$ denote the experimental bounds of $3\sigma$ (global analyses) and $2\sigma$ (T2K) ranges, respectively:
		(a)  $\delta_{CP}$ versus $\sin^2\theta_{23}$,
		(b)  $\delta_{CP}$ versus Majorana phase $\beta$,
		(c)  $\delta_{CP}$ versus $k$,
		and (d)   $\delta_{CP}$ versus $\phi_k$.}
	\label{fig:13NH}
\end{figure}

\begin{figure}[h!]
	\begin{tabular}{cc}
		\begin{minipage}{0.5\hsize}
			\begin{center}
				\includegraphics[width=\linewidth]{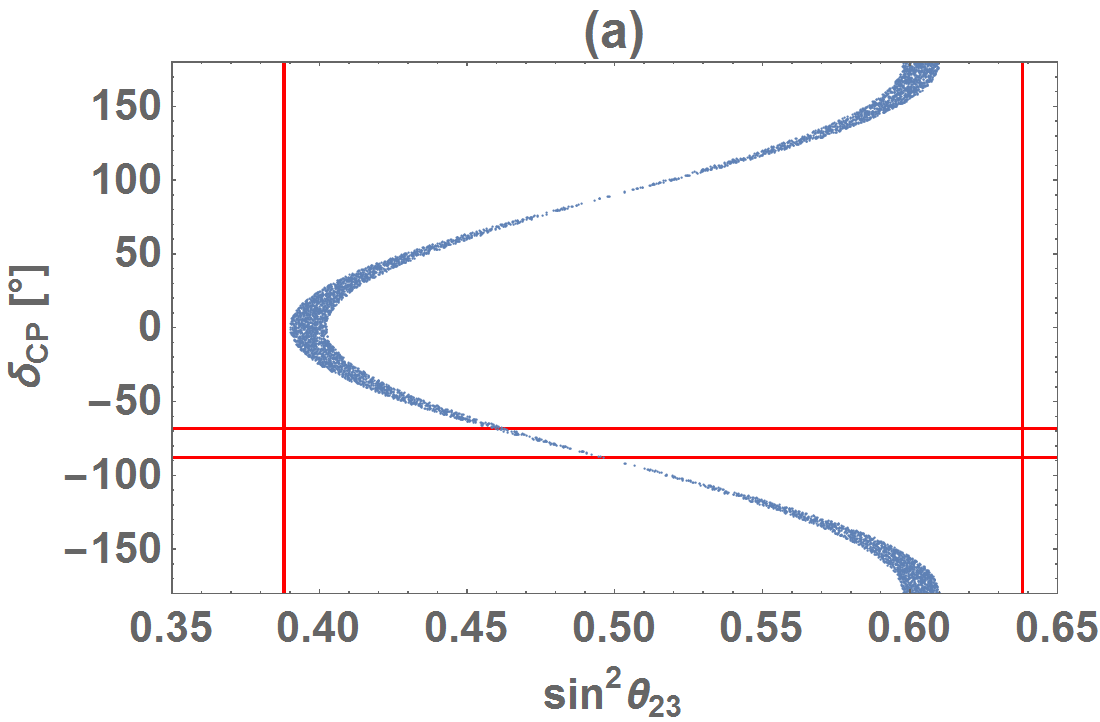}
			\end{center}
		\end{minipage}
		\begin{minipage}{0.5\hsize}
			\begin{center}
				\includegraphics[width=\linewidth]{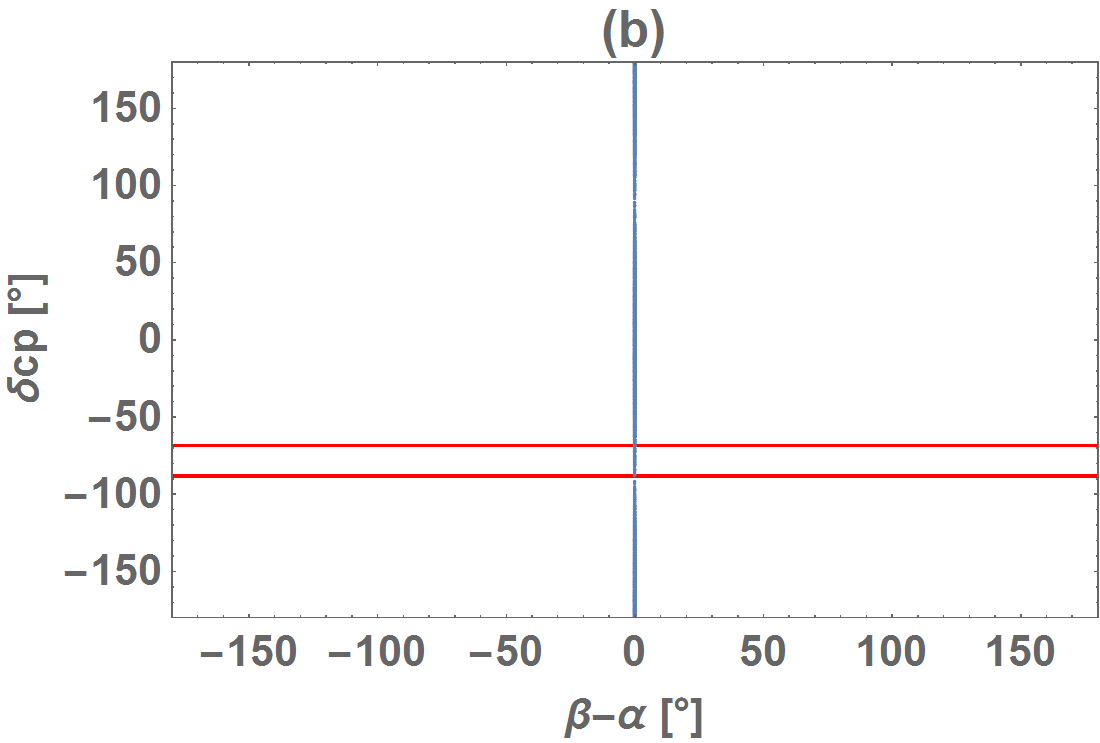}
			\end{center}
		\end{minipage}
	\end{tabular}
	\begin{tabular}{cc}
		\begin{minipage}{0.5\hsize}
			\begin{center}
				\includegraphics[width=\linewidth]{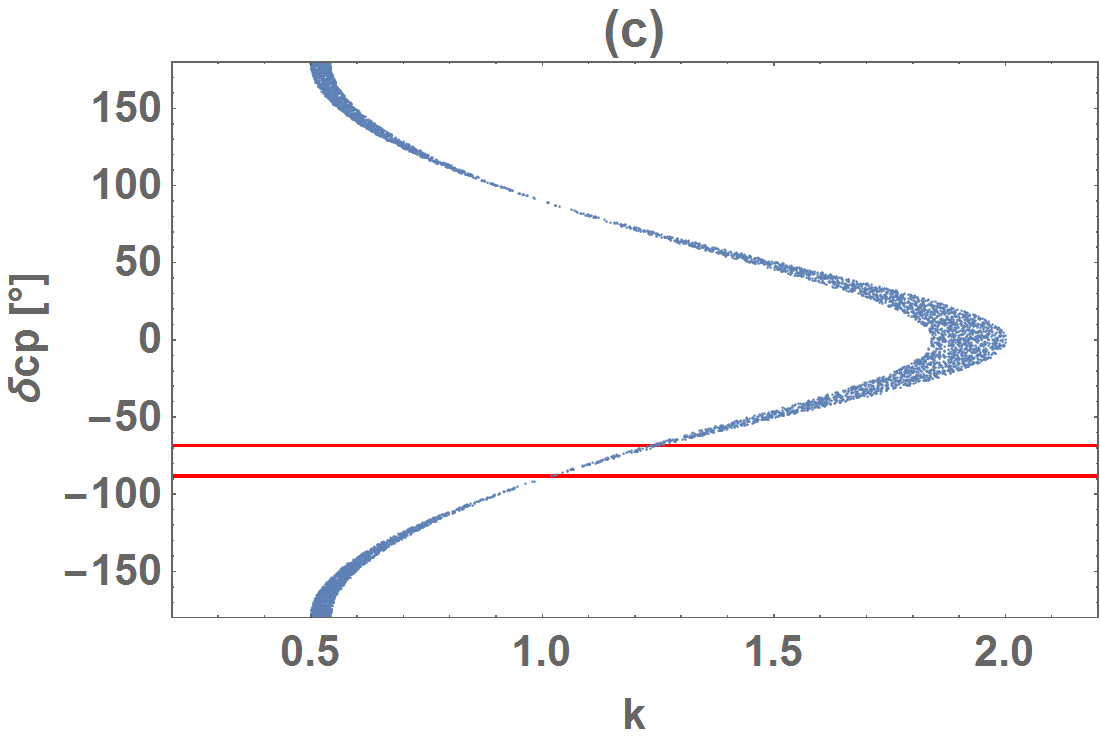}
			\end{center}
		\end{minipage}
		\begin{minipage}{0.5\hsize}
			\begin{center}
				\includegraphics[width=\linewidth]{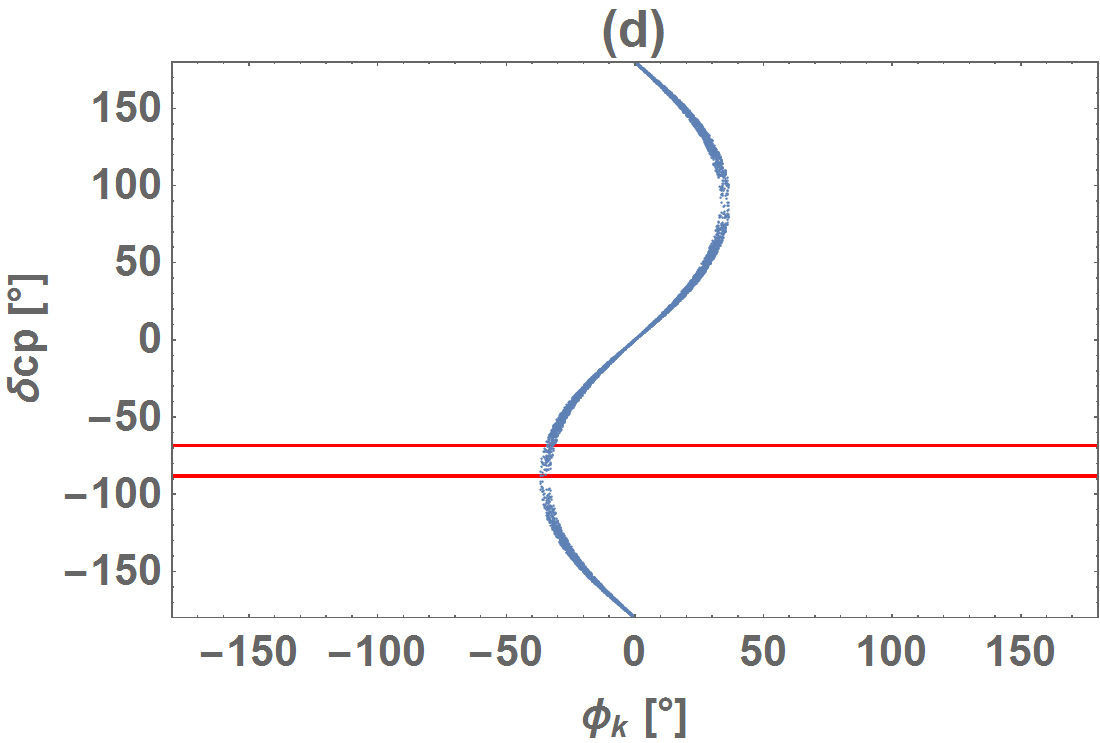}
			\end{center}
		\end{minipage}
	\end{tabular}
	\caption{Predictions for $\rm TM_2$ in IH,
		 where
	the	red lines for $\sin^2\theta_{23}$ and $\delta_{CP}$ denote the experimental bounds of $3\sigma$ (global analyses) and $2\sigma$ (T2K), respectively:
		(a)  $\delta_{CP}$ versus $\sin^2\theta_{23}$,
		(b)  $\delta_{CP}$ versus difference of two Majorana phases $\beta-\alpha$,
		(c)  $\delta_{CP}$ versus $k$,
		and (d)   $\delta_{CP}$ versus $\phi_k$.}
	\label{fig:13IH}
\end{figure}

In Figs.~\ref{fig:13NH} and \ref{fig:13IH}, we show the numerical results for NH and IH, respectively.
For both NH and IH cases, $\delta_{CP}$ is allowed to take all range in $(-\pi,\pi)$ as seen in Figs.~\ref{fig:13NH}(a) and~\ref{fig:13IH}(a).
It is found that the Dirac CP violating phase becomes maximal $\pm \pi/2$ around the maximal mixing angle of $\theta_{23}=\pi/4$.
For NH, the Majorana phase $\beta$ is allowed to take all values from $-\pi$ to $\pi$ as seen in Fig.~\ref{fig:13NH}(b).
On the other hand, for IH, the difference of two Majorana phases $\beta-\alpha$ is almost zero as seen in Fig.~\ref{fig:13IH}(b).

The predicted value of $\delta_{CP}$ is sensitive to the parameter $k$ which behaves differently for NH and IH.
In the case of NH,
$\delta_{CP}$ is $0$ or $\pm\pi$ at the lower bound of  $k$, $0.80$
or at the upper bound, $1.25$, respectively, as seen in Fig.~\ref{fig:13NH}(c).
In the region of $k=0.80\sim 1.25$, the predicted  $\delta_{CP}$
is changed drastically.
On the other hand, for the case of IH,
$\delta_{CP}$ is $0$ or $\pm\pi$ at the lower bound $k=0.50$
or at the upper bound $k=2.00$, respectively, as seen in Fig.~\ref{fig:13IH}(c).
In the region of $k=0.50\sim 2.00$,  $\delta_{CP}$ is
predicted for each $k$.
The phase $\phi_k$ is restricted in the narrow region of $\pm (165^\circ \sim 180^\circ )$ for NH as seen in Fig.~\ref{fig:13NH}(d).
On the other hand, for IH,
the phase $\phi_k$ is in the  region of $-37^\circ\sim 37^\circ$ as seen in Fig.~\ref{fig:13IH}(d).
In both cases,
 $\delta_{CP}$ and $\theta_{23}$ become maximal simultaneously if $k=1$ is taken.

The effective mass $|m_{ee}|$ is predicted in the region around $1.6\sim 4.2$ meV and $50$ meV for NH and IH, respectively.

\section{Summary and discussions}
We have studied  the minimal seesaw model
where only two right-handed Majorana neutrinos are assumed,
focusing on the CP violating phase $\delta_{CP}$.
In addition, we have taken the  trimaximal  mixing pattern  for the  neutrino flavor ($\rm TM_1$ or $\rm TM_2$) where the  charged lepton mass matrix is diagonal.
Owing to this  symmetric  framework,
the flavor structure of the $3\times 2$ Dirac neutrino mass matrix is given in terms of a few parameters.

We have  examined  three cases of the Dirac neutrino mass matrix for $\rm TM_1$ in NH.
It is emphasized that
the observation of the CP violating phase determines the flavor structure of the Dirac neutrino mass matrix in the minimal seesaw model.
New minimal  Dirac neutrino mass matrices have been presented as the result of the numerical study for case I.
Our model includes the Littlest seesaw model by King {\it et al.}
as the  one of the specific cases.
We have also discussed the case of $\rm TM_1$ in IH.
The structure of the Dirac neutrino mass matrix is restricted considerably.
The parameters are determined  in the narrow  ranges,
 $k=0.65\sim 1.40$ and $\phi_k=\pm (25^\circ \sim 38^\circ )$.
 We have  studied the case of $\rm TM_2$ in  NH and IH.
The predicted  $\delta_{CP}$ is allowed to take all range in $(-\pi,\pi)$, but
it is sensitive  to  the parameter $k$, which is around $1$.
In both cases,
 $\delta_{CP}$ and the mixing angle $\theta_{23}$ become maximal simultaneously for  $k=1$.

Our $3\times 2$ Dirac neutrino mass matrix is reproduced by introducing gauge singlet flavons with the VEV's in  $S_4$ flavor symmetry.
The specific alignments of the VEV's are derived from the residual symmetry of $S_4$ group.
It is interesting to consider the underlying mechanism which dynamically realizes the required vacuum alignments.

Finally, we add comments.
 If our Dirac neutrino mass matrices are given at the high energy scale, for example, the GUT scale,
 one should examine the renormalization group correction for the neutrino mixing matrix.
 However, it is very small since the lightest neutrino mass vanishes in our framework
as seen in Ref.~\cite{Gehrlein:2016fms}.

The sign of  the CP violating  phase $\delta_{CP}$ is not determined in our framework.
As well known,
 CP violating phases in the neutrino mass matrix are related to CP violating phases at the high energy.
Since one can discuss the baryon asymmetry of the universe assuming the leptogenesis \cite{Fukugita:1986hr},
the sign of $\delta_{CP}$ can be predicted  in our framework.
This work will appear elsewhere.

\vspace{0.4 cm}
\noindent
{\bf Acknowledgment}
   
We thank Akihiro Yu for careful reading of the manuscript. 
This work is supported by JSPS Grants-in-Aid for Scientific Research
 16J05332 (YS) and 15K05045, 16H00862 (MT).

\appendix

\section*{Appendix}


\section{Minimal seesaw mass matrix}

We can take the $2\times 2$ right-handed Majorana neutrino mass matrix $M_R$ to be real diagonal in general:
\begin{equation}
M_R=-M_0
\begin{pmatrix}
p^{-1} & 0 \\
0 & 1
\end{pmatrix},
\end{equation}
where $M_0$ is the mass scale of the right-handed Majorana neutrino and $p$ is the ratio between the two right-handed Majorana neutrino masses.
The minus sign in front of $M_0$ is taken as our sign convention.
On the other hand, the relevant Dirac neutrino mass matrix $M_D$ is defined as
\begin{equation}
M_D=
\begin{pmatrix}
a & d \\
b & e \\
c & f
\end{pmatrix},
\end{equation}
where $a\sim f$ are complex parameters.
By using the seesaw mechanism, the left-handed Majorana neutrino mass matrix $M_{\nu }$ is given by
\begin{equation}
M_{\nu }=M_DM_R^{-1}M_D^T=\frac{1}{M_0}
\begin{pmatrix}
a^2p+d^2 & abp+de & acp+df \\
abp+de & b^2p+e^2 & bcp+ef \\
acp+df & bcp+ef & c^2p+f^2
\end{pmatrix}.
\label{Aleft-handed-Majorana-1}
\end{equation}
By turning the neutrino mass matrix $M_\nu$ to the TBM mixing basis, $M_{\nu }$ is given as
\begin{align}
\hat M_\nu\equiv  V_{\text{TBM}}^T M_\nu V_{\text{TBM}}=\frac{1}{M_0}
\begin{pmatrix}
\frac{A_\nu ^2p+D_\nu^2}{6} & \frac{A_\nu B_\nu p+D_\nu  E_\nu}{3\sqrt{2}} &
\frac{A_\nu C_\nu p+D_\nu F_\nu}{2\sqrt{3}} \\
\frac{A_\nu B_\nu p+D_\nu  E_\nu}{3\sqrt{2}} & \frac{B_\nu^2p+E_\nu^2}{3} &
\frac{B_\nu C_\nu p+E_\nu F_\nu}{\sqrt{6}} \\
\frac{A_\nu C_\nu p+D_\nu F_\nu}{2\sqrt{3}} &\frac{B_\nu C_\nu p+E_\nu F_\nu}{\sqrt{6}} & \frac{C_\nu^2p+F_\nu^2}{2}
\end{pmatrix},
\label{Aleft-handed-Majorana-TBM}
\end{align}
where
\begin{align}
&& A_\nu\equiv 2a-b-c, \qquad B_\nu\equiv a+b+c, \qquad C_\nu\equiv c-b, \nonumber\\
&& D_\nu\equiv 2d-e-f, \qquad E_\nu\equiv d+e+f, \qquad F_\nu\equiv f-e.
\end{align}
We discuss neutrino mass structures to realize the additional 2-3 family rotation ($\rm TM_1$) and the additional 1-3 one ($\rm TM_2$)
to the TBM mixing basis  for both  NH and IH.

\subsection{$\rm TM_1$: Additional 2-3 family  rotation in NH}
At first, we consider the case of NH in $\rm TM_1$.
Since $(1,1)$, $(1,2)$, $(2,1)$, $(1,3)$, and $(3,1)$ entries of the matrix must be zero,
conditions for the additional 2-3 rotation to the TBM mixing are
\begin{equation}
A_\nu=2a-b-c=0,\qquad D_\nu=2d-e-f=0,
\end{equation}
where $(a,b,c)$ are supposed to be independent of $(d,e,f)$.
After imposing these conditions on the Eq.~(\ref{Aleft-handed-Majorana-TBM}), the mass matrix is rewritten as
\begin{equation}
\hat M_\nu =\frac{1}{M_0}
\begin{pmatrix}
0 & 0 & 0 \\
0 & \frac{3}{4}\left ((b+c)^2p+(e+f)^2\right ) &
\frac{1}{2}\sqrt{\frac{3}{2}}\left ((c^2-b^2)p-e^2+f^2\right ) \\
0 & \frac{1}{2}\sqrt{\frac{3}{2}}\left ((c^2-b^2)p-e^2+f^2\right ) &
\frac{1}{2}\left ((b-c)^2p+(e-f)^2\right )
\end{pmatrix},
\label{Aleft-handed-Majorana-general}
\end{equation}
where the lightest neutrino mass $m_1$ is zero.
Since the Majorana neutrino mass can be rescaled in the seesaw formula,
the right-handed Majorana and Dirac neutrino mass matrices are written by putting $p=1$ as
\begin{equation}
M_R=-M_0
\begin{pmatrix}
1 & 0 \\
0 & 1
\end{pmatrix},\qquad
M_D=
\begin{pmatrix}
\frac{b+c}{2} & \frac{e+f}{2} \\
b & e \\
c & f
\end{pmatrix},
\label{Ageneral-texture}
\end{equation}
respectively.
Starting from these textures, we discuss the specific cases which are attractive in the standpoint of the flavor model.

We consider one zero textures leading to the additional 2-3 family rotation to the TBM mixing basis.
There are three possible patterns of one zero texture as follows:
\begin{equation}
\mbox{(I) $b+c=0$,\qquad (I\hspace{-.1em}I) $c=0$, \qquad
	(I\hspace{-.1em}I\hspace{-.1em}I) $b=0$,}
\end{equation}
in  Eq.~(\ref{Ageneral-texture}).
Corresponding Dirac neutrino mass matrices can be obtained as
\begin{equation}
M_D=\left\{
\begin{array}{cl}
\begin{pmatrix}
0 & \frac{e+f}{2} \\
b & e \\
-b & f
\end{pmatrix} & \mbox{ for (I) $b+c=0$}\\
\begin{pmatrix}
\frac{b}{2} & \frac{e+f}{2} \\
b & e \\
0 & f
\end{pmatrix} & \mbox{ for (I\hspace{-.1em}I) $c=0$}\\
\begin{pmatrix}
\frac{c}{2} & \frac{e+f}{2} \\
0 & e \\
c & f
\end{pmatrix} & \mbox{ for (I\hspace{-.1em}I\hspace{-.1em}I) $b=0$}
\end{array}
\right . .\label{AMD1}
\end{equation}
One can get another set by exchanging between the first column and second one in the Dirac neutrino mass matrix of Eq.~(\ref{AMD1}).
However, the neutrino mass matrix $\hat M_\nu$ is invariant by this exchange.
Therefore, we consider only three cases in  Eq.~(\ref{AMD1}).

We show the neutrino mass matrix $\hat M_\nu$ for three cases:
\begin{equation}
{\rm Case\ I}\ : \
\hat M_{\nu }=\frac{1}{M_0}
\begin{pmatrix}
0 & 0 & 0 \\
0 & \frac{3}{4}(e+f)^2 & -\frac{1}{2}\sqrt{\frac{3}{2}}(e-f)(e+f) \\
0 & -\frac{1}{2}\sqrt{\frac{3}{2}}(e-f)(e+f) & 2b^2+\frac{1}{2}(e-f)^2
\end{pmatrix}
,
\end{equation}

\begin{equation}
{\rm Case\ II}\ : \
\hat M_{\nu }=\frac{1}{M_0}
\begin{pmatrix}
0 & 0 & 0 \\
0 & \frac{3}{4}[b^2+(e+f)^2] & -\frac{1}{2}\sqrt{\frac{3}{2}}[b^2+(e-f)(e+f)] \\
0 &  -\frac{1}{2}\sqrt{\frac{3}{2}}[b^2+(e-f)(e+f)] & \frac{1}{2}[b^2+(e-f)^2]
\end{pmatrix}
,
\end{equation}

\begin{equation}
{\rm Case\ III}\ : \
\hat M_{\nu }=\frac{1}{M_0}
\begin{pmatrix}
0 & 0 & 0 \\
0 & \frac{3}{4}[c^2+(e+f)^2] & -\frac{1}{2}\sqrt{\frac{3}{2}}[-c^2+(e-f)(e+f)] \\
0 &  -\frac{1}{2}\sqrt{\frac{3}{2}}[-c^2+(e-f)(e+f)] & \frac{1}{2}[c^2+(e-f)^2]
\end{pmatrix}
.
\end{equation}

\subsection{$\rm TM_1$: Additional 2-3 rotation in IH}
Let us discuss the case of IH in $\rm TM_1$.
In order to give the additional 2-3 family rotation to the TBM mixing,
the $(1,2)$, $(1,3)$, $(2,1)$, and $(3,1)$ elements in Eq.~(\ref{Aleft-handed-Majorana-TBM}) should vanish.
These conditions are given as
\begin{equation}
A_\nu=a+b+c=0,\qquad C_\nu=c-b=0,\qquad D_\nu=2d-e-f=0.
\end{equation}
By setting $p=1$, we have
\begin{equation}
M_R=-M_0
\begin{pmatrix}
1 & 0 \\
0 & 1
\end{pmatrix}, \quad \quad M_D=
\begin{pmatrix}
-2b & \frac{e+f}{2} \\
b & e \\
b& f
\end{pmatrix}.
\label{Ageneral-texture-inverted}
\end{equation}
The neutrino mass matrix $\hat M_\nu$ is given as
\begin{equation}
\hat M_{\nu }=\frac{1}{M_0}
\begin{pmatrix}
6b^2 & 0 & 0 \\
0 & \frac{3}{4}(e+f)^2 & -\frac{1}{2}\sqrt{\frac{3}{2}}(e-f)(e+f) \\
0 & -\frac{1}{2}\sqrt{\frac{3}{2}}(e-f)(e+f) & \frac{1}{2}(e-f)^2
\end{pmatrix}
,
\end{equation}
where the neutrino mass $m_3$ vanishes.
\subsection{$\rm TM_2$: Additional 1-3 rotation in NH or IH}

Let us consider the case of the additional 1-3 family rotation to the  TBM mixing basis.
This case is called as $\rm TM_2$.
Then, $(1,2)$, $(2,3)$, $(2,1)$, and $(3,2)$ elements in Eq.~(\ref{Aleft-handed-Majorana-TBM}) should vanish.
These conditions are given as
\begin{equation}
A_\nu=2a-b-c=0, \qquad C_\nu=c-b=0, \qquad E_\nu=d+e+f=0\ .
\end{equation}
The neutrino mass matrix $\hat M_\nu$ is written by
\begin{equation}
\hat M_\nu =\frac{1}{M_0}
\begin{pmatrix}
\frac{3}{2}(e+f)^2 & 0 &\frac{\sqrt{3}}{2}(e^2-f^2)\\
0 & 3b^2 & 0 \\
\frac{\sqrt{3}}{2}(e^2-f^2) &0 & \frac{1}{2}(e-f)^2
\end{pmatrix},
\label{Aleft-handed-Majorana-general-inverted}
\end{equation}
where $p=1$ is set.
The right-handed Majorana  and the Dirac neutrino mass matrices are
\begin{equation}
M_R=-M_0
\begin{pmatrix}
1 & 0 \\
0 & 1
\end{pmatrix}, \qquad
 M_D=
\begin{pmatrix}
b & -e-f \\
b & e \\
b& f
\end{pmatrix},
\label{Ageneral-texture-inverted}
\end{equation}
respectively.
There is another solution
\begin{equation}
B_\nu=a+b+c=0, \quad D_\nu=2d-e-f=0, \quad F_\nu=f-e=0\ .
\end{equation}
However, this set leads to same structure of the neutrino mass matrix as in Eq.~(\ref{Aleft-handed-Majorana-general-inverted}).
The mass eigenvalue $m_1$ or $m_3$ vanishes for NH or IH, respectively.


\section{Representations of $S_4$ group}
\label{sec:S4}
We show the representations of $S_4$ group in this appendix.
All elements of $S_4$ group are expressed by products of the generators $s$ and $t$, which satisfy
\begin{equation}
s^4=t^3=e,\quad st^2s=t,\quad sts=ts^2t,
\end{equation}
where $e$ is an identity element.
These generators are represented on ${\bf 1_1}$, ${\bf 1_2}$, ${\bf 2}$, ${\bf 3_1}$, and ${\bf 3_2}$
of $S_4$ group as follows~\cite{Ishimori:2010au,Ishimori:2012zz}:
\begin{align}
{\bf 1_1}&:\quad s=1,\quad t=1, \nonumber \\
{\bf 1_2}&:\quad s=-1,\quad t=1, \nonumber \\
{\bf 2}&:\quad s=
\begin{pmatrix}
0 & 1 \\
1 & 0
\end{pmatrix},\quad t=
\begin{pmatrix}
\omega & 0 \\
0 & \omega ^2
\end{pmatrix}, \nonumber \\
{\bf 3_1}&:\quad s=\frac{1}{3}
\begin{pmatrix}
-1 & 2\omega  & 2\omega ^2 \\
2\omega & 2\omega ^2 & -1 \\
2\omega ^2 & -1 & 2\omega
\end{pmatrix},\quad t=
\begin{pmatrix}
1 & 0 & 0 \\
0 & \omega ^2 & 0 \\
0 & 0 & \omega
\end{pmatrix}, \nonumber \\
{\bf 3_2}&:\quad s=-\frac{1}{3}
\begin{pmatrix}
-1 & 2\omega  & 2\omega ^2 \\
2\omega & 2\omega ^2 & -1 \\
2\omega ^2 & -1 & 2\omega
\end{pmatrix},\quad t=
\begin{pmatrix}
1 & 0 & 0 \\
0 & \omega ^2 & 0 \\
0 & 0 & \omega
\end{pmatrix}.
\label{ishimori}
\end{align}
On the other hand, in Ref.~\cite{Hagedorn:2010th,King:2013eh}, all elements of $S_4$ group are also expressed by products of the three generators $S$, $T$, and $U$, which satisfy
\begin{equation}
S^2=T^3=U^2=(ST)^3=(SU)^2=(TU)^2=(STU)^4=e.
\end{equation}
Note that the minimal number of $S_4$ generators is only two.
However, in order to compare generators of $S_4$ group with that of $A_4$ group,
it is convenient to express elements of the group in terms of $S$, $T$, and $U$.
These generators are also represented on
${\bf 1}$, ${\bf 1'}$, ${\bf 2}$, ${\bf 3}$, and ${\bf 3'}$
of $S_4$ group as follows:
\begin{align}
{\bf 1}&:\quad S=1,\quad T=1,\quad U=1, \nonumber \\
{\bf 1'}&:\quad S=1,\quad T=1,\quad U=-1, \nonumber \\
{\bf 2}&:\quad S=
\begin{pmatrix}
1 & 0 \\
0 & 1
\end{pmatrix},\quad T=
\begin{pmatrix}
\omega & 0 \\
0 & \omega ^2
\end{pmatrix},\quad U=
\begin{pmatrix}
0 & 1 \\
1 & 0
\end{pmatrix}, \nonumber \\
{\bf 3}&:\quad S=\frac{1}{3}
\begin{pmatrix}
-1 & 2 & 2 \\
2 & -1 & 2 \\
2 & 2 & -1
\end{pmatrix},\quad T=
\begin{pmatrix}
1 & 0 & 0 \\
0 & \omega ^2 & 0 \\
0 & 0 & \omega
\end{pmatrix},\quad U=-
\begin{pmatrix}
1 & 0 & 0 \\
0 & 0 & 1 \\
0 & 1 & 0
\end{pmatrix}, \nonumber \\
{\bf 3'}&:\quad S=\frac{1}{3}
\begin{pmatrix}
-1 & 2 & 2 \\
2 & -1 & 2 \\
2 & 2 & -1
\end{pmatrix},\quad T=
\begin{pmatrix}
1 & 0 & 0 \\
0 & \omega ^2 & 0 \\
0 & 0 & \omega
\end{pmatrix},\quad U=
\begin{pmatrix}
1 & 0 & 0 \\
0 & 0 & 1 \\
0 & 1 & 0
\end{pmatrix},
\end{align}
where ${\bf 1}$ corresponds to ${\bf 1_1}$,
${\bf 1'}$ corresponds to ${\bf 1_2}$, ${\bf 3}$ corresponds to ${\bf 3_2}$, and
${\bf 3'}$ corresponds to ${\bf 3_1}$ in Eq.~(\ref{ishimori}), respectively.
It is remarked that
$s$ and $t$ generators are related with $S$, $T$, and $U$ as follows:
\begin{equation}
s=STUST \ ,\qquad t=T \ .
\end{equation}
The tensor products of $S_4$ group are independent of the basis of representations and written as follows:
\begin{align}
{\bf 1}\otimes {\bf r}&={\bf r}\otimes {\bf 1}={\bf r},\qquad {\bf 1'}\otimes {\bf 1'}={\bf 1},\qquad
{\bf 1'}\otimes {\bf 2}={\bf 2}\otimes {\bf 1'}={\bf 2}, \nonumber \\
{\bf 1'}\otimes {\bf 3}&={\bf 3}\otimes {\bf 1'}={\bf 3'},\qquad
{\bf 1'}\otimes {\bf 3'}={\bf 3'}\otimes {\bf 1'}={\bf 3}, \nonumber \\
{\bf 2}\otimes {\bf 2}&={\bf 1}\oplus {\bf 1'}\oplus {\bf 2},\qquad
{\bf 2}\otimes {\bf 3}={\bf 3}\otimes {\bf 2}={\bf 2}\otimes {\bf 3'}={\bf 3'}\otimes {\bf 2}=
{\bf 3}\oplus {\bf 3'}, \nonumber \\
{\bf 3}\otimes {\bf 3}&={\bf 3'}\otimes {\bf 3'}={\bf 1}\oplus {\bf 2}\oplus {\bf 3}\oplus {\bf 3'},\qquad
{\bf 3}\otimes {\bf 3'}={\bf 3'}\otimes {\bf 3}={\bf 1'}\oplus {\bf 2}\oplus {\bf 3}\oplus {\bf 3'},
\end{align}
where ${\bf r}$ is an arbitrary representation of $S_4$ group.

For the following tensor products,
\begin{equation}
{\bf 3}\otimes {\bf 3} \     \rightarrow {\bf 1}, \qquad
{\bf 3'}\otimes {\bf 3'} \rightarrow{\bf 1}, \qquad
{\bf 3}\otimes {\bf 3'}\ \rightarrow {\bf 1'} ,
\end{equation}
the Clebsch-Gordan coefficient is given as
\begin{equation}
\alpha_1 \beta_1 + \alpha_2 \beta_3 + \alpha_3 \beta_2  \  ,
\end{equation}
where $\alpha_i$ and $\beta_i$ are elements of $\bf 3^{(')}$.

\end{document}